\documentclass{article}

    \PassOptionsToPackage{numbers, compress}{natbib}

    \usepackage[final]{neurips_2025}

\usepackage[utf8]{inputenc} 
\usepackage[T1]{fontenc}    
\usepackage{hyperref}       
\usepackage{url}            
\usepackage{booktabs}       
\usepackage{amsfonts}       
\usepackage{nicefrac}       
\usepackage{microtype}      
\usepackage{graphicx}
\usepackage{wrapfig}
\usepackage{adjustbox}
\usepackage{subfigure}
\usepackage[ruled]{algorithm2e}
\usepackage{bbm}
\usepackage{epsfig}
\usepackage{amsmath,bm}
\usepackage{amssymb}
\usepackage{bbding}
\usepackage{color}
\usepackage{colortbl}
\usepackage[dvipsnames,table,xcdraw]{xcolor}
\usepackage{tabularx}
\usepackage[normalem]{ulem}
\useunder{\uline}{\ul}{}
\usepackage{multirow}
\usepackage{xurl}

\usepackage{breakurl} 
\hypersetup{breaklinks=true}
\usepackage{ragged2e}
\usepackage{rotating}
\usepackage{tablefootnote}
\usepackage{makecell}
\usepackage{threeparttable} 
\usepackage{pifont}
\usepackage{hyperref}
\usepackage{algpseudocode}
\usepackage{amsmath}
\usepackage{url}
\usepackage{booktabs}
\usepackage{breakurl} 
\hypersetup{breaklinks=true}
\usepackage{xurl}
\usepackage{multirow}
\usepackage{colortbl} 
\usepackage[table,xcdraw]{xcolor}
\usepackage[normalem]{ulem}
\useunder{\uline}{\ul}{}
\usepackage{tabularx}
\usepackage{ragged2e}
\usepackage{rotating}
\usepackage{tablefootnote}
\usepackage{makecell}
\usepackage{threeparttable} 
\usepackage{pifont}
\usepackage[dvipsnames]{xcolor}
\definecolor{headerblue}{RGB}{173,216,230} 
\definecolor{lightgray}{gray}{0.95} 
\usepackage[most]{tcolorbox}
\usepackage{float}
\usepackage{booktabs}
\usepackage{colortbl}
\usepackage{array}
\usepackage[table]{xcolor}
\newcolumntype{C}[1]{>{\centering\arraybackslash}m{#1}}
\title{ChromFound: Towards A Universal Foundation Model for Single-Cell Chromatin Accessibility Data}

\author{%
    \textbf{Yifeng Jiao\textsuperscript{1,2,\thanks{Equal contribution.}}},
    \textbf{Yuchen Liu\textsuperscript{1,2,\footnotemark[1]}}, 
    \textbf{Yu Zhang\textsuperscript{2}},
    \textbf{Xin Guo\textsuperscript{1,2,\thanks{Corresponding author.}}},
    \textbf{Yushuai Wu\textsuperscript{2}},
    \textbf{Chen Jiang\textsuperscript{1,2}},\\
    \textbf{Jiyang Li\textsuperscript{2}},
    \textbf{Hongwei Zhang\textsuperscript{1,2}},
    \textbf{Limei Han\textsuperscript{1,2}},
    \textbf{Xin Gao\textsuperscript{2,3,4,5}},
    \textbf{Yuan Qi\textsuperscript{1,2,6,\footnotemark[2]}},
    \textbf{Yuan Cheng\textsuperscript{1,2,\footnotemark[2]}}
    \\
    \textsuperscript{1}Artificial Intelligence Innovation and Incubation Institute, Fudan University \\
    \textsuperscript{2}Shanghai Academy of Artificial Intelligence for Science \\
    \textsuperscript{3}Computer, Electrical and Mathematical Sciences and Engineering Division, KAUST \\
    \textsuperscript{4}Center of Excellence for Smart Health, KAUST \\
    \textsuperscript{5}Center of Excellence on GenAI, KAUST \\
    \textsuperscript{6}Zhongshan Hospital, Fudan University \\
    Correspondence to: \{jiaoyifeng,guoxin\}@sais.org.cn, \{cheng\_yuan,qiyuan\}@fudan.edu.cn
}

\begin{document}

\maketitle

\begin{abstract}
The advent of single-cell Assay for Transposase-Accessible Chromatin using sequencing (scATAC-seq) offers an innovative perspective for deciphering regulatory mechanisms by assembling a vast repository of single-cell chromatin accessibility data. While foundation models have achieved significant success in single-cell transcriptomics, there is currently no foundation model for scATAC-seq that supports zero-shot high-quality cell identification and comprehensive multi-omics analysis simultaneously. Key challenges lie in the high dimensionality and sparsity of scATAC-seq data, as well as the lack of a standardized schema for representing open chromatin regions (OCRs). Here, we present \textbf{ChromFound}, a foundation model tailored for scATAC-seq. ChromFound utilizes a hybrid architecture and genome-aware tokenization to effectively capture genome-wide long contexts and regulatory signals from dynamic chromatin landscapes. Pretrained on 1.97 million cells from 30 tissues and 6 disease conditions, ChromFound demonstrates broad applicability across 6 diverse tasks. Notably, it achieves robust zero-shot performance in generating universal cell representations and exhibits excellent transferability in cell type annotation and cross-omics prediction. By uncovering enhancer-gene links undetected by existing computational methods, ChromFound offers a promising framework for understanding disease risk variants in the noncoding genome. The implementation of ChromFound is available via https://github.com/JohnsonKlose/ChromFound.
\end{abstract}

\section{Introduction}
\label{introduction}
The human genome harbors an extensive repository of open chromatin regions (OCRs) responsible for orchestrating precise spatial and temporal gene expression patterns~\cite{Klemm2019ChromatinAA}. Identifying OCRs at single-cell level is pivotal for understanding the regulatory landscape of individual cells, offering profound insights into cellular diversity and dynamic mechanisms of gene regulation~\cite{cusanovich2018cis, corces2020single}. The single-cell Assay for Transposase-Accessible Chromatin using sequencing (scATAC-seq)~\cite{buenrostro2015single} stands as the most prevalent genome-wide technique to enable the identification of OCRs at single-cell resolution. Its principal applications span cancer research, immunological studies, and neuroscience~\cite{wang2022single,terekhanova2023epigenetic,sundaram2024single,yang2024pan}. Furthermore, with initiatives such as the Human Cell Atlas~\cite{rood2024human} and other significant studies~\cite{domcke2020human,zhang2021single}, the scale of scATAC-seq data is expanding rapidly, opening new possibilities and presenting novel challenges in the exploration of OCRs at the same time.


Pretrained language models (PLMs) have achieved great success in deciphering the complex language of single-cell RNA sequencing (scRNA-seq) data~\cite{geneformer,cui2024scgpt,hao2024large,yang2022scbert}, validating their capabilities of fine-tuning for a variety of downstream tasks.
Nevertheless, the development of generalizable foundation models for scATAC-seq remains under-explored, posing significant challenges for cellular representation learning and multi-task transfer. Here, we identify and discuss three major challenges.


(1). \textbf{scATAC-seq data are inherently high-dimensional and sparse}, with single cells spanning over millions of OCRs but typically less than 1\% showing accessibility. This sparsity complicates the utilization of Transformer~\cite{vaswani2017attention} adopted in transcriptomics~\cite{geneformer,cui2024scgpt,hao2024large,wen2023cellplm,fu2025foundation,yang2022scbert}. Other methods such as transforming accessibility into gene activity~\cite{granja2021archr, signac}, filtering highly variable OCRs~\cite{granja2021archr,tayyebi2024scalable,ghazanfar2024stabilized}, or binarizing accessibility~\cite{hao2021integrated, cusanovich2015multiplex,yuan2022scbasset} always cause substantial information loss~\cite{su2025scmultimap,martens2024modeling,chen2019assessment}.

(2). \textbf{scATAC-seq data are not formatted in a standardized feature space}. Dynamic chromatin landscapes across diverse sources and platforms make dictionary-based tokenization for genes and nucleotides unsuitable for OCRs. All deep learning models for scATAC-seq~\cite{xiong2019scale,xiong2022online,cui2024discrete,yuan2022scbasset} require dataset-specific training, leading to poor zero-shot performance.


(3). \textbf{current methods lack a unified integration of genomic information with chromatin accessibility profiles}. Both genomic information defining regulatory loci and chromatin accessibility profiles revealing their activity are essential in scATAC-seq data. However, VAE-based models~\cite{xiong2019scale,xiong2022online,cui2024discrete} process accessibility matrices independently, while models like scBasset, CellSpace, and SANGO~\cite{yuan2022scbasset,tayyebi2024scalable,zeng2024deciphering} predict binary accessibility from genomic sequences. These methods struggle to generalize across diverse downstream tasks.


To address these challenges, we present \textbf{ChromFound}, a universal foundation model for single-cell chromatin accessibility data. Its hybrid architecture integrates a Mamba block for efficient long-range sequence processing with a self-attention block to capture local regulatory dependencies within ±200 kb of transcriptional start sites, aligning with the typical range of enhancer-promoter interactions identified by Hi-C analysis~\cite{Bonev2016OrganizationAF}. ChromFound incorporates biologically informed OCR tokenization encoding genomic coordinates and non-binary accessibility profiles. This component ensures scalability and alignment across diverse scATAC-seq datasets from varied tissues, platforms, and protocols. Pretrained on over 1.97 million single cells from more than 30 organs or tissues and 6 major disease categories~(e.g., Alzheimer’s, Parkinson’s, leukemia, glioma), ChromFound leverages 1.86 trillion training tokens, surpassing the scale of existing models such as Geneformer~\cite{geneformer}, scGPT~\cite{cui2024scgpt}, and scFoundation~\cite{hao2024large}, all of which use fewer than 1 trillion tokens.

To validate the effectiveness of ChromFound, we conduct comprehensive evaluations across multiple tasks and datasets. In cell representation, ChromFound outperforms baselines instead of additional training, achieving an average ARI improvement of 17.02\% across 8 scATAC-seq datasets from 4 tissues. Additional results demonstrate its robustness to denoise technical effect in zero-shot settings. For cell type annotation, ChromFound significantly improves accuracy and macro F1 score across all evaluated datasets and tissues. In cross-omics prediction, it accurately infers gene expression from chromatin accessibility, surpassing all baselines across 5 datasets. Furthermore, in the K562 cell line, ChromFound effectively identifies gene-enhancer relationships and regulatory perturbation effects, highlighting its potential for interpretable \textit{cis}-regulatory modeling.

Our main contributions can be summarized as follows:

(1) ChromFound, the first scATAC-seq foundation model, employs a genome-wide architecture along with a genome-aware tokenization and is pretrained on a large corpus with 1.97 million single cells.

(2) ChromFound provides a strong zero-shot cell representation from an epigenetic perspective, as well as showing consistent robustness to data sparsity and batch effect.

(3) ChromFound shows great transfer learning capabilities in cell annotation and cross-omics prediction, validating biological significance in inferring enhancer-gene links and perturbation responses.

\section{Related Work}
Single-cell foundation models for scRNA-seq data have seen significant advancements, supporting tasks such as gene function prediction, cell annotation, and drug response modeling. Geneformer~\cite{geneformer}, scGPT~\cite{cui2024scgpt}, and scFoundation~\cite{hao2024large} are trained by large-scale datasets, while scBERT~\cite{yang2022scbert} and CellPLM~\cite{wen2023cellplm} focus on specific objectives based on pre-training language models such as imputation and cell communications. However, existing models are predominantly tailored for scRNA-seq data and fail to model scATAC-seq data effectively. GET~\cite{fu2025foundation} achieves a level of predictive precision comparable to experimental replicates. It does so by integrating chromatin accessibility data and genomic sequence information in a pseudo-bulk format, rather than relying on single-cell resolution.

Deep learning has been widely applied to scATAC-seq data analysis for cell clustering and annotation. PeakVI~\cite{peakvi} and PoissonVI~\cite{poissonvi} extend VAE-based frameworks, with the former disentangling biological from technical variation for batch correction and the latter introducing a Poisson likelihood to preserve quantitative accessibility. SCALE~\cite{xiong2019scale} and SCALEX~\cite{xiong2022online} employ VAEs for latent representation learning and multi-omics integration, while CASTLE~\cite{cui2024discrete} enhances cell-type annotation via self-supervised learning. Beyond VAE-based methods, SnapATAC2~\cite{snapatac} scales to millions of cells using latent semantic indexing and spectral embedding, and Signac~\cite{signac} provides a Seurat-based workflow for dimensionality reduction and multimodal integration. scBasset~\cite{yuan2022scbasset} links genomic sequences with accessibility profiles, and Cellcano~\cite{ma2023cellcano} together with SANGO~\cite{zeng2024deciphering} employ two-step MLPs for cell-type annotation. Despite their advances, these models remain limited by architectural constraints and data scarcity, restricting scalability and generalization.

\begin{figure}
  \centering
  \includegraphics[width=\textwidth]{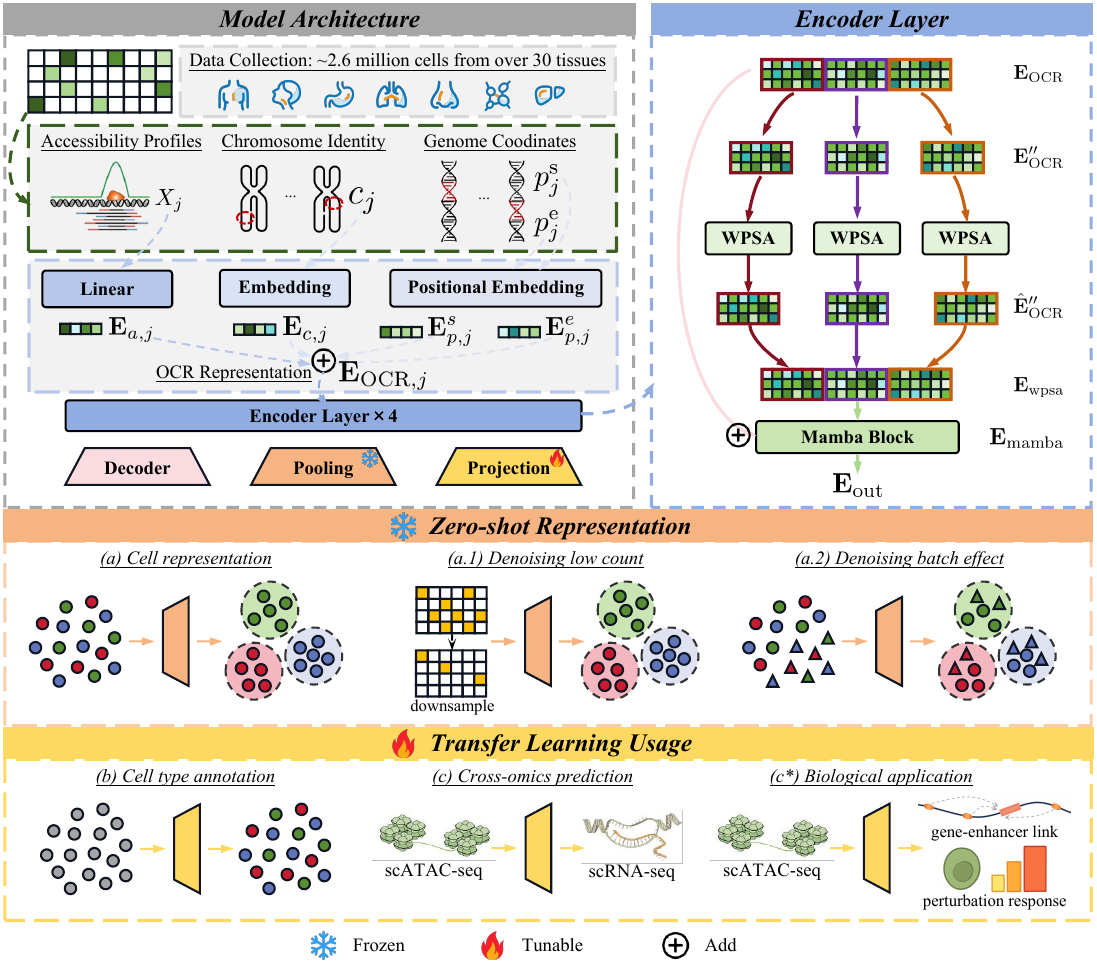}
  \caption{\textbf{An overview of ChromFound architecture}. The methods of ChromFound are listed in Section~\ref{sec:methods}. OCR tokenization are described in Section~\ref{tokenization}. The hybrid encoder layer is described in Section~\ref{sec:encoder}. Other pretraining details of datasets and implementation are included in Section~\ref{sec:pretrainig}. All evaluation results are detailed in Section~\ref{sec:experiments}.}
  \label{architecture}
\end{figure}

\section{Methods}
\label{sec:methods}
\subsection{OCR Tokenization}
\label{tokenization}
To encode chromatin accessibility data, we treat each OCR as a token by three essential components: chromosome embedding, positional embedding for genomic coordinates, and continuous accessibility embedding. The tokenization integrates these components to form a comprehensive representation of each OCR. Our unified token representation is designed to address two fundamental challenges. First, by encoding start and end genomic coordinates through sinusoidal positional embeddings, ChromFound captures the full extent of each OCR, accommodating varying fragment lengths. Second, by avoiding reliance on a fixed OCR vocabulary, ChromFound flexibly represents tissue-specific or novel OCRs, thereby mitigating misalignment and dropout across tissues. In the following, we describe the encoding strategy for each component in detail.

\textbf{Chromosome Embedding}: Chromosome identity is encoded using a learnable embedding lookup table. The size of the chromosome embedding lookup table is 25, with 24 corresponding to the number of human chromosomes and 1 designated for the padding token. We define the learnable embedding matrix $\mathbf{W}_{c} \in \mathbb{R}^{|\mathcal{C}| \times d_{\text{model}}}$ and the chromosome index $c_{j}$ of OCR $j$. ChromFound maps the $c_{j}$ to $\mathbf{E}_{c, j}$ by the following equation:
\begin{equation}
    \mathbf{E}_{c, j} = \mathbf{W}_{c}[c_{j}, :] .
\end{equation}
\textbf{Positional Embedding for Genomic Coordinates}: Since some studies~\cite{liu2024mamba4rec, xu2024sstmultiscalehybridmambatransformer} have noted that state space models inherently handle sequential information through their recurrent nature, we use positional embedding to encode genomic coordinates instead of the sequence order. Specifically, the embedding of the starting position and the ending position for the OCR $j$ is computed as:

\begin{equation}
\mathbf{E}^{s}_{p, j} = 
\begin{bmatrix}
\sin\left(\frac{p^{s}_{j}}{\text{temp}\cdot10000^{2k / d_{\text{model}}}}\right) \\
\cos\left(\frac{p^{s}_{j}}{\text{temp}\cdot10000^{2k / d_{\text{model}}}}\right)
\end{bmatrix}, \quad
\mathbf{E}^{e}_{p, j} = 
\begin{bmatrix}
\sin\left(\frac{p^{e}_{j}}{\text{temp}\cdot10000^{2k / d_{\text{model}}}}\right) \\
\cos\left(\frac{p^{e}_{j}}{\text{temp}\cdot10000^{2k / d_{\text{model}}}}\right)
\end{bmatrix}
\end{equation}
where $p^{s}_{j}$ and $p^{e}_{j}$ represent the starting and end position for the OCR $j$ on the GRCh38 reference genome, $\text{temp}$ is a hyperparameter set to 100000 in ChromFound, and $d_{\text{model}}$ is the dimension of the embedding  layer. The index $k$ spans $\{0, 1, \dots, d_{\text{model}}/2\}$, with sine and cosine functions alternately assigned to even and odd dimensions of the embedding vector.

\textbf{Accessibility Embedding}: The accessibility value for the $j$th OCR, denoted as $X_{j}$, is a continuous scalar processed by the procedures detailed in Appendix~\ref{sec:data_processing}. To project the value $X_{j}$ into the hidden dimension, we use a linear transformation as $\mathbf{E}_{a, j} = \text{Linear}(X_{j})$.

\textbf{OCR Representation}: The unified representation \(\mathbf{E}_{\text{OCR}, j} \in \mathbb{R}^{1 \times D} \) of the $j$th OCR is obtained by adding the previous embeddings together:
\begin{equation}
\mathbf{E}_{\text{OCR}, j} = \mathbf{E}_{c, j} + \mathbf{E}^{s}_{p, j} + \mathbf{E}^{e}_{p, j} + \mathbf{E}_{a, j}.
\end{equation}

\subsection{Hybrid Encoder Layer}
\label{sec:encoder}

The ChromFound encoder layer transforms the OCR representation~\( \mathbf{E}_\text{OCR} \in \mathbb{R}^{L \times D} \) into a latent space, where \(L\) denotes the sequence length and \(D\) is the embedding dimension. As illustrated in Fig.~\ref{architecture}, ChromFound incorporates a self-attention module~(Section~\ref{sec:wpsa}) to effectively capture local dependencies among OCRs, and integrates a Mamba layer~(Section~\ref{sec:mamba}) to handle ultra-long OCR sequences. This architecture is motivated by biological insights, recognizing that both short- and long-range interactions among OCRs are critical to gene regulatory mechanisms~\cite{kim2015architectural,panigrahi2021mechanisms,ramasamy2023mediator}.

\subsubsection{Window Partition Self-Attention (WPSA)}
\label{sec:wpsa}

Due to its computational complexity $O(L^2)$, the vanilla attention mechanism~\cite{vaswani2017attention} is not directly applicable to scATAC-seq data containing millions of OCRs. To address these challenges, ChromFound introduces an optimized self-attention layer named \textbf{W}indow \textbf{P}artition \textbf{S}elf-\textbf{A}ttention (WPSA), which applies the attention mechanism within local windows of size \( W \), where \( W \ll L \).

\textbf{Window Partition}: Given the OCR representation \( \mathbf{E}_\text{OCR} \in \mathbb{R}^{L \times D} \), \(L\) may not be a multiple of \(W\). We define the number of windows $ N = \left\lceil \frac{L}{W} \right\rceil $ and pad \(\mathbf{E}_\text{OCR}\) with zeros up to \(N \times W\) along the dimension \(L\). As a result, the padded sequence \( \mathbf{E}_\text{OCR}' \in \mathbb{R}^{(N \cdot W) \times D} \) is reshaped as \( \mathbf{E}''_{\text{OCR}} \in \mathbb{R}^{N \times W \times D} \).

\textbf{Self-Attention}: In practice, we set the window size to \(W=256\), which approximately spans 200 kb upstream and downstream of transcriptional start site (TSS). This setting is consistent with the scope of enhancer-promoter interactions suggested by high-resolution Hi-C analysis~\cite{Bonev2016OrganizationAF}.

\textbf{Window Restore}: After applying self-attention within each window, we merge outputs of each window $\hat{\mathbf{E}}''_{\text{OCR}}$ back into \(\hat{\mathbf{E}}_{\text{wpsa}}\):
\begin{equation}
\hat{\mathbf{E}}_{\text{wpsa}} = \mathrm{concat}\bigl[\hat{\mathbf{E}}''_{\text{OCR},1},\ldots,\hat{\mathbf{E}}''_{\text{OCR},N}] \;\in\; \mathbb{R}^{(N \cdot W) \times D}.
\end{equation}
Furthermore, we remove all zero padding tokens in $\hat{\mathbf{E}}_{\text{wpsa}}$ and get the WPSA output $\mathbf{E}_{\text{wpsa}} \in \mathbb{R}^{L \times D}$. 

\subsubsection{Mamba Block}
\label{sec:mamba}
Mamba~\cite{gu2023mamba} is capable of processing extended sequences due to the linear-time complexity of state space models (SSMs)~\cite{gu2021combining, gu2021efficiently}, making it particularly well suited for modeling high-resolution scATAC-seq data. When combined with the WPSA module, which amplifies local signals within predefined windows, the Mamba block enables effective integration of both short- and long-range dependencies among OCRs across the genome. This hybrid architecture mitigates the degradation of long-range information typically caused by relying solely on zero-order discretization~\cite{orvieto2023resurrecting,yu2024robustifying}. To reduce parameter count and refine representations, we project $\mathbf{E}_{\text{wpsa}}$ to a lower-dimensional space $\mathbb{R}^{L \times D_{\mathrm{low}}}$ and then back, which is defined as:
\begin{equation}
\mathbf{E}_{\mathrm{down}} = \mathbf{E}_{\text{wpsa}} W_{\mathrm{down}}, \quad \mathbf{E}_{\mathrm{mamba}} = \mathrm{Mamba}(\mathbf{E}_{\mathrm{down}}), \quad \mathbf{E}_{\mathrm{up}} = \mathbf{E}_{\mathrm{mamba}} W_{\mathrm{up}},
\end{equation}
where $W_{\mathrm{down}} \in \mathbb{R}^{D \times D_{\mathrm{low}}}$ ($D_{\mathrm{low}} < D$) and $W_{\mathrm{up}} \in \mathbb{R}^{D_{\mathrm{low}} \times D}$.

\subsubsection{Overall Workflow}
The overall workflow of the encoder layer is illustrated in Fig.~\ref{architecture} and proceeded as:
\begin{equation}
\mathbf{E}_{\text{out}} = \mathbf{E}_{\text{OCR}} + W_{\mathrm{up}} \cdot \text{Mamba}(W_{\mathrm{down}} \cdot \text{WPSA}(\text{RMSNorm}(\mathbf{E}_{\text{OCR}})))
\end{equation}

\subsection{Pre-training}
\label{sec:pretrainig}
\textbf{Datasets}:~We assemble a large-scale scATAC-seq dataset comprising over 2.64 million cells detailed in Table~\ref{tab:dataset_info}. For model pretraining, we select a representative subset of 1.97 million cells from more than 30 distinct organs and tissues. To rigorously evaluate the generalization ability of ChromFound, we additionally compile a benchmark dataset containing 0.67 million cells from diverse sources, including tissues such as bone and retina excluded from pretraining. 

\textbf{Implementation}:~ChromFound is trained for 5 epochs over 80 hours on a compute cluster comprising 4 machines with a total of 32 NVIDIA A100 GPUs. The effective batch size is set to 128. We use the AdamW optimizer~\cite{loshchilov2017decoupled} with a maximum learning rate of $5 \times 10^{-5}$. The embedding dimension and the hidden size of WPSA \(D\) are set to 128, while dimension \(D_{\text{low}}\) of the Mamba block is set to 32. The model architecture consists of 4 stacked encoder layers in total.

\textbf{Training objective}:~Due to the high sparsity of scATAC-seq data, ChromFound adopts a masking strategy that simultaneously targets both zero and non-zero OCR values. Inspired by xTrimoGene~\cite{gong2024xtrimogene}, the setting encourages the model to predict non-binary chromatin accessibility across the entire profile, mitigating the risk of learning representations dominated by zero entries. Poisson/ZINB losses often cause unstable optimization due to gradients dominated by nonzero entries. Instead, mean squared error (MSE) loss on log-transformed, normalized signals ensures stability. The MLP layer is employed as the decoder to reconstruct the original input \(X_{i}\). The reconstruction output of cell $i$ is denoted as \(\hat{X}_{i}\). The pretraining objective is the MSE loss computed over the masked positions \(\mathcal{M}_{i}\) in each cell $i$:
\begin{equation}
    \mathcal{L}_{mse,i} = \frac{1}{|\mathcal{M}_{i}|} \sum_{i \in \mathcal{M}_{i}} \left( X_{i} - \hat{X}_{i} \right)^2.
\end{equation}

\begin{table}
\caption{Results of cell clustering tasks. Result style: \textbf{best}, \uline{second best}, \textcolor{OliveGreen}{\textbf{relative gains}}.}
\renewcommand{\arraystretch}{0.68} 
\scriptsize
\resizebox{\textwidth}{!}{%
\begin{tabular}{@{}c|c|c|cccc@{}}
\toprule
\textbf{Dataset}                                                                          & \textbf{Tissue}         & \textbf{Model}      & \textbf{ARI}($\uparrow$)  & \textbf{FMI}($\uparrow$)  & \textbf{NMI}($\uparrow$)  & \textbf{AMI}($\uparrow$)  \\ \midrule
\multirow{6}{*}[-4.5ex]{\begin{tabular}[c]{@{}c@{}}Morabito130K~\cite{cortex63k}\\ (batch 1)\end{tabular}}         & \multirow{6}{*}[-4.5ex]{Cortex} & PeakVI~\cite{peakvi}              & 0.4558±0.0038 & 0.6286±0.0022 & 0.6964±0.0012 & 0.6940±0.0012 \\
                                                                                          &                         & PoissonVI~\cite{poissonvi}              & 0.3673±0.0028 & 0.5673±0.0018 & 0.5878±0.0001 & 0.5845±0.0001 \\
                                                                                          &                         & SnapATAC2~\cite{snapatac}              & 0.5118±0.0315 & 0.6740±0.0145 & 0.7202±0.0041 & 0.7179±0.0042 \\
                                                                                          &                         & Signac~\cite{signac}              & 0.5013±0.0507 & \uline{0.7035±0.0177} & 0.5580±0.0160 & 0.5534±0.0165 \\
                                                                                          &                         & SCALE~\cite{xiong2019scale}               & \uline{0.5460±0.0074} & 0.6896±0.0042 & \uline{0.7282±0.0016} & \uline{0.7275±0.0016} \\
                                                                                          &                         & SCALEX~\cite{xiong2022online}              & 0.4930±0.0022 & 0.6505±0.0013 & 0.6958±0.0001 & 0.6949±0.0001 \\
                                                                                          &                         & CASTLE~\cite{cui2024discrete}              & 0.4425±0.0029 & 0.6109±0.0019 & 0.7086±0.0012 & 0.7077±0.0012 \\
                                                                                          &                         & scBasset~\cite{yuan2022scbasset}            & 0.5039±0.0039 & 0.6747±0.0022 & 0.6674±0.0005 & 0.6647±0.0005 \\
                                                                                          &                         & \textbf{ChromFound} & \textbf{0.6890±0.0387} & \textbf{0.7943±0.0144} & \textbf{0.7779±0.0037} & \textbf{0.7760±0.0038} \\ \cline{3-7}\rule{0pt}{2.5ex}
                                                                                          &                         & \textbf{Gains(\%)}  & \textcolor{OliveGreen}{\textbf{26.20}}         & \textcolor{OliveGreen}{\textbf{15.17}}         & \textcolor{OliveGreen}{\textbf{6.81}}          & \textcolor{OliveGreen}{\textbf{6.66}}          \\ \midrule
\multirow{6}{*}[-4.5ex]{\begin{tabular}[c]{@{}c@{}}Morabito130K~\cite{cortex63k}\\ (batch 2)\end{tabular}}         & \multirow{6}{*}[-4.5ex]{Cortex} & PeakVI~\cite{peakvi}               & 0.5512±0.0016 & 0.6394±0.0032 & \uline{0.7050±0.0006} & \uline{0.7028±0.0006} \\
                                                                                          &                         & PoissonVI~\cite{poissonvi}               & 0.5334±0.0052 & 0.6388±0.0033 & 0.6004±0.0001 & 0.5983±0.0001 \\
                                                                                          &                         & SnapATAC2~\cite{snapatac}               & 0.5325±0.0072 & 0.6380±0.0045 & 0.7038±0.0018 & 0.7015±0.0018 \\
                                                                                          &                         & Signac~\cite{signac}               & 0.5069±0.0123 & \uline{0.6478±0.0064} & 0.5888±0.0032 & 0.5851±0.0033 \\
                                                                                          &                         & SCALE~\cite{xiong2019scale}               & 0.5306±0.0048 & 0.6394±0.0032 & 0.6221±0.0007 & 0.6200±0.0007 \\
                                                                                          &                         & SCALEX~\cite{xiong2022online}              & 0.3905±0.0041 & 0.5252±0.0027 & 0.5361±0.0028 & 0.5336±0.0029 \\
                                                                                          &                         & CASTLE~\cite{cui2024discrete}              & 0.3743±0.0016 & 0.5087±0.0010 & 0.4636±0.0021 & 0.4608±0.0022 \\
                                                                                          &                         & scBasset~\cite{yuan2022scbasset}            & \uline{0.5373±0.0024} & 0.6424±0.0015 & 0.6659±0.0003 & 0.6641±0.0003 \\
                                                                                          &                         & \textbf{ChromFound} & \textbf{0.6278±0.0043} & \textbf{0.7156±0.0025} & \textbf{0.7321±0.0016} & \textbf{0.7299±0.0016} \\ \cline{3-7}\rule{0pt}{2.5ex} 
                                                                                          &                         & \textbf{Gains(\%)}  & \textcolor{OliveGreen}{\textbf{16.83}}         & \textcolor{OliveGreen}{\textbf{10.47}}         & \textcolor{OliveGreen}{\textbf{3.84}}          & \textcolor{OliveGreen}{\textbf{3.86}}          \\ \midrule
\multirow{6}{*}[-4.5ex]{\begin{tabular}[c]{@{}c@{}}Kuppe139K~\cite{kuppe2022spatial}\\ (donar av3)\end{tabular}}          & \multirow{6}{*}[-4.5ex]{Heart} & PeakVI~\cite{peakvi}               & 0.4159±0.0007 & 0.5434±0.0006 & 0.6024±0.0005 & 0.6000±0.0006 \\
                                                                                          &                         & PoissonVI~\cite{poissonvi}               & 0.3969±0.0005 & 0.5232±0.0004 & 0.5717±0.0001 & 0.5688±0.0001 \\
                                                                                          &                         & SnapATAC2~\cite{snapatac}               & \uline{0.4571±0.0099} & 0.5564±0.0064 & 
                                                                                          \uline{0.6625±0.0013} & 0.6502±0.0013 \\
                                                                                          &                         & Signac~\cite{signac}               & 0.3123±0.0047 & 0.4601±0.0028 & 0.5272±0.0028 & 0.5238±0.0029 \\
                                                                                          &                         & SCALE~\cite{xiong2019scale}               & 0.4475±0.0015 & 0.5654±0.0011 & 0.6005±0.0003 & 0.5978±0.0003 \\
                                                                                          &                         & SCALEX~\cite{xiong2022online}              & 0.3819±0.0007 & 0.5061±0.0005 & 0.4829±0.0000 & 0.4797±0.0000 \\
                                                                                          &                         & CASTLE~\cite{cui2024discrete}              & 0.3886±0.0017 & 0.5118±0.0110 & 0.4765±0.0001 & 0.4729±0.0001 \\
                                                                                          &                         & scBasset~\cite{yuan2022scbasset}            & 0.4553±0.0022 & \uline{0.5729±0.0016} & 0.6539±0.0006 & \uline{0.6516±0.0006} \\
                                                                                          &                         & \textbf{ChromFound} & \textbf{0.5828±0.0052} & \textbf{0.6757±0.0034} & \textbf{0.7207±0.0008} & \textbf{0.7187±0.0008} \\ \cline{3-7}\rule{0pt}{2.5ex} 
                                                                                          &                         & \textbf{Gains(\%)}  & \textcolor{OliveGreen}{\textbf{27.75}}         & \textcolor{OliveGreen}{\textbf{17.94}}         & \textcolor{OliveGreen}{\textbf{8.78}}         & \textcolor{OliveGreen}{\textbf{10.31}}         \\ \midrule
\multirow{6}{*}[-4.5ex]{\begin{tabular}[c]{@{}c@{}}Kuppe139K~\cite{kuppe2022spatial}\\ (donar av10)\end{tabular}}         & \multirow{6}{*}[-4.5ex]{Heart} & PeakVI~\cite{peakvi}               & 0.4301±0.0030 & 0.6383±0.0021 & 0.6149±0.0022 & 0.6144±0.0022 \\
                                                                                          &                         & PoissonVI~\cite{poissonvi}               & 0.3606±0.0054 & 0.5929±0.0027 & 0.4472±0.0066 & 0.4464±0.0066 \\
                                                                                          &                         & SnapATAC2~\cite{snapatac}               & 0.5515±0.0312 & 0.7285±0.0124 & \uline{0.6867±0.0060} & \uline{0.6863±0.0060} \\
                                                                                          &                         & Signac~\cite{signac}               & 0.1159±0.0112 & 0.5093±0.0053 & 0.2294±0.0076 & 0.2281±0.0076 \\
                                                                                          &                         & SCALE~\cite{xiong2019scale}               & 0.4517±0.0087 & 0.6596±0.0040 & 0.5397±0.0041 & 0.5391±0.0041 \\
                                                                                          &                         & SCALEX~\cite{xiong2022online}              & 0.4742±0.0105 & 0.6770±0.0055 & 0.5040±0.0079 & 0.5033±0.0080 \\
                                                                                          &                         & CASTLE~\cite{cui2024discrete}              & 0.4739±0.0000 & 0.6786±0.0000 & 0.5204±0.0000 & 0.5197±0.0000 \\
                                                                                          &                         & scBasset~\cite{yuan2022scbasset}            & \uline{0.5823±0.0111} & \uline{0.7459±0.0050} & 0.6780±0.0001 & 0.6776±0.0001 \\
                                                                                          &                         & \textbf{ChromFound} & \textbf{0.6774±0.0445} & \textbf{0.8109±0.0169} & \textbf{0.7369±0.0137} & \textbf{0.7365±0.0138} \\ \cline{3-7}\rule{0pt}{2.5ex} 
                                                                                          &                         & \textbf{Gains(\%)}  & \textcolor{OliveGreen}{\textbf{16.34}}         & \textcolor{OliveGreen}{\textbf{8.71}}          & \textcolor{OliveGreen}{\textbf{7.31}}          & \textcolor{OliveGreen}{\textbf{7.31}}          \\ \midrule
\multirow{6}{*}[-4.5ex]{\begin{tabular}[c]{@{}c@{}}Liang154K~\cite{liang2023multi}\\ (sample D026\_13)\end{tabular}}    & \multirow{6}{*}[-4.5ex]{Retina} & PeakVI~\cite{peakvi}               & 0.2885±0.0024 & 0.5543±0.0019 & 0.5644±0.0019 & 0.5638±0.0019 \\
                                                                                          &                         & PoissonVI~\cite{poissonvi}               & 0.3162±0.0025 & 0.5766±0.0017 & 0.6081±0.0017 & 0.6076±0.0017 \\
                                                                                          &                         & SnapATAC2~\cite{snapatac}               & \uline{0.5606±0.0569} & 0.7207±0.0199 & 0.7045±0.0085 & 0.7041±0.0085 \\
                                                                                          &                         & Signac~\cite{signac}               & 0.5538±0.1051 & \uline{0.7451±0.0095} & 0.4931±0.0535 & 0.4919±0.0538 \\
                                                                                          &                         & SCALE~\cite{xiong2019scale}               & 0.5512±0.0386 & 0.7285±0.0075 & \uline{0.7172±0.0051} & \uline{0.7168±0.0051} \\
                                                                                          &                         & SCALEX~\cite{xiong2022online}              & 0.5069±0.0048 & 0.7241±0.0022 & 0.6759±0.0005 & 0.6755±0.0005 \\
                                                                                          &                         & CASTLE~\cite{cui2024discrete}              & 0.2424±0.0032 & 0.5136±0.0022 & 0.5298±0.0033 & 0.5292±0.0033 \\
                                                                                          &                         & scBasset~\cite{yuan2022scbasset}            & 0.4330±0.0049 & 0.6699±0.0029 & 0.6903±0.0015 & 0.6898±0.0015 \\
                                                                                          &                         & \textbf{ChromFound} & \textbf{0.6668±0.0500} & \textbf{0.8149±0.0169} & \textbf{0.7644±0.0093} & \textbf{0.7641±0.0093} \\ \cline{3-7}\rule{0pt}{2.5ex} 
                                                                                          &                         & \textbf{Gains(\%)}  & \textcolor{OliveGreen}{\textbf{18.94}}         & \textcolor{OliveGreen}{\textbf{9.37}}         & \textcolor{OliveGreen}{\textbf{6.59}}          & \textcolor{OliveGreen}{\textbf{6.60}}          \\ \midrule
\multirow{6}{*}[-4.5ex]{\begin{tabular}[c]{@{}c@{}}Liang154K~\cite{liang2023multi}\\ (sample D19D008)\end{tabular}}     & \multirow{6}{*}[-4.5ex]{Retina} & PeakVI~\cite{peakvi}               & 0.4702±0.0061 & 0.6318±0.0037 & 0.7229±0.0024 & 0.7224±0.0025 \\
                                                                                          &                         & PoissonVI~\cite{poissonvi}               & 0.4978±0.0122 & 0.6527±0.0065 & 0.7400±0.0028 & 0.7395±0.0028 \\
                                                                                          &                         & SnapATAC2~\cite{snapatac}               & 0.5814±0.0019 & 0.7173±0.0010 & 0.7881±0.0011 & 0.7877±0.0011 \\
                                                                                          &                         & Signac~\cite{signac}               & 0.5146±0.0532 & 0.7339±0.0119 & 0.5235±0.0307 & 0.5222±0.0309 \\
                                                                                          &                         & SCALE~\cite{xiong2019scale}               & \uline{0.6129±0.0131} & \uline{0.7465±0.0113} & \uline{0.7981±0.0020} & \uline{0.7993±0.0058} \\
                                                                                          &                         & SCALEX~\cite{xiong2022online}              & 0.5424±0.0041 & 0.6872±0.0023 & 0.6976±0.0019 & 0.6971±0.0019 \\
                                                                                          &                         & CASTLE~\cite{cui2024discrete}              & 0.4272±0.0052 & 0.5982±0.0034 & 0.6614±0.0034 & 0.6608±0.0034 \\
                                                                                          &                         & scBasset~\cite{yuan2022scbasset}            & 0.6027±0.0139 & 0.7313±0.0066 & 0.7881±0.0010 & 0.7878±0.0010 \\
                                                                                          &                         & \textbf{ChromFound} & \textbf{0.6688±0.0264} & \textbf{0.7767±0.0123} & \textbf{0.8183±0.0032} & \textbf{0.8179±0.0032} \\ \cline{3-7}\rule{0pt}{2.5ex} 
                                                                                          &                         & \textbf{Gains(\%)}  & \textcolor{OliveGreen}{\textbf{9.12}}          & \textcolor{OliveGreen}{\textbf{4.05}}          & \textcolor{OliveGreen}{\textbf{2.53}}          & \textcolor{OliveGreen}{\textbf{2.34}}          \\ \midrule
\multirow{6}{*}[-4.5ex]{\begin{tabular}[c]{@{}c@{}}PBMC169K~\cite{de2024systematic}\\ (batch VIB\_10xv1\_1)\end{tabular}} & \multirow{6}{*}[-4.5ex]{PBMC}   & PeakVI~\cite{peakvi}               & 0.5417±0.0022 & 0.6340±0.0015 & 0.6639±0.0009 & 0.6627±0.0009 \\
                                                                                          &                         & PoissonVI~\cite{poissonvi}               & \uline{0.6314±0.0027} & 0.7073±0.0018 & 0.7353±0.0012 & 0.7343±0.0012 \\
                                                                                          &                         & SnapATAC2~\cite{snapatac}               & 0.6246±0.0112 & \uline{0.7085±0.0067} & 0.7551±0.0023 & 0.7541±0.0023 \\
                                                                                          &                         & Signac~\cite{signac}               & 0.1832±0.0136 & 0.4543±0.0043 & 0.3064±0.0198 & 0.3027±0.0201 \\
                                                                                          &                         & SCALE~\cite{xiong2019scale}               & 0.6158±0.0048 & 0.6959±0.0031 & 0.7632±0.0010 & 0.7623±0.0010 \\
                                                                                          &                         & SCALEX~\cite{xiong2022online}              & 0.6121±0.0048 & 0.6927±0.0015 & 0.7593±0.0008 & 0.7584±0.0008 \\
                                                                                          &                         & CASTLE~\cite{cui2024discrete}              & 0.5530±0.0018 & 0.6439±0.0012 & 0.7263±0.0004 & 0.7252±0.0004 \\
                                                                                          &                         & scBasset~\cite{yuan2022scbasset}            & 0.6240±0.0001 & 0.7019±0.0001 & \uline{0.7660±0.0000} & \uline{0.7651±0.0000} \\
                                                                                          &                         & \textbf{ChromFound} & \textbf{0.6953±0.0035} & \textbf{0.7601±0.0023} & \textbf{0.7860±0.0007} & \textbf{0.7852±0.0008} \\ \cline{3-7}\rule{0pt}{2.5ex} 
                                                                                          &                         & \textbf{Gains(\%)}  & \textcolor{OliveGreen}{\textbf{10.12}}         & \textcolor{OliveGreen}{\textbf{7.28}}          & \textcolor{OliveGreen}{\textbf{2.61}}          & \textcolor{OliveGreen}{\textbf{2.62}}          \\ \midrule
\multirow{6}{*}[-4.5ex]{\begin{tabular}[c]{@{}c@{}}PBMC169K~\cite{de2024systematic}\\ (batch BIO\_ddseq\_1)\end{tabular}} & \multirow{6}{*}[-4.5ex]{PBMC}   & PeakVI~\cite{peakvi}               & 0.2915±0.0005 & 0.5041±0.0004 & 0.4623±0.0004 & 0.4615±0.0005 \\
                                                                                          &                         & PoissonVI~\cite{poissonvi}               & 0.4103±0.0045 & 0.5994±0.0028 & 0.5656±0.0035 & 0.5649±0.0035 \\
                                                                                          &                         & SnapATAC2~\cite{snapatac}               & 0.4126±0.0318 & 0.6149±0.0149 & \uline{0.5761±0.0071} & 
                                                                                          \uline{0.5754±0.0071} \\
                                                                                          &                         & Signac~\cite{signac}               & 0.1885±0.0097 & 0.5249±0.0056 & 0.2852±0.0040 & 0.2839±0.0040 \\
                                                                                          &                         & SCALE~\cite{xiong2019scale}               & 0.4359±0.0181 & \uline{0.6250±0.0101} & 0.5594±0.0045 & 0.5587±0.0045 \\
                                                                                          &                         & SCALEX~\cite{xiong2022online}              & 0.4087±0.0197 & 0.6063±0.0106 & 0.5519±0.0066 & 0.5512±0.0066 \\
                                                                                          &                         & CASTLE~\cite{cui2024discrete}              & 0.3504±0.0042 & 0.5555±0.0026 & 0.5512±0.0013 & 0.5506±0.0013 \\
                                                                                          &                         & scBasset~\cite{yuan2022scbasset}            & \uline{0.4362±0.0012} & 0.6223±0.0008 & 0.5577±0.0008 & 0.5571±0.0008 \\
                                                                                          &                         & \textbf{ChromFound} & \textbf{0.4835±0.0082} & \textbf{0.6604±0.0042} & \textbf{0.5950±0.0036} & \textbf{0.5944±0.0036} \\ \cline{3-7}\rule{0pt}{2.5ex} 
                                                                                          &                         & \multicolumn{1}{c|}{\textbf{Gains(\%)}}  & \textcolor{OliveGreen}{\textbf{10.84}}         & \textcolor{OliveGreen}{\textbf{5.66}}          & \textcolor{OliveGreen}{\textbf{3.28}}          & \textcolor{OliveGreen}{\textbf{3.30}}          \\ \bottomrule
\end{tabular}%
\label{clustering_table}
}
\end{table}

\section{Experiments}
\label{sec:experiments}
To systematically evaluate the performance of ChromFound, we conduct experiments on various tissues and datasets. We first assess its zero-shot representation capabilities, demonstrating strong robustness to both data sparsity and batch effects. We then evaluate the transferability in downstream tasks, including cell type annotation and cross-omics prediction. Finally, we validate the biological utility of ChromFound by accurately inferring gene-enhancer regulatory links and gene expression responses to enhancer perturbations. All scATAC-seq datasets used in experiments are preprocessed according to the procedures detailed in Appendix~\ref{sec:data_processing}.

\subsection{Cell representation}
\label{sec:clustering}
ChromFound generates low-dimensional cell representations by applying PCA to the encoder outputs after the pooling layer. Details of the implementations are provided in Appendix~\ref{sec:cell_clustering_add}. For comparison, we include scBasset~\cite{yuan2022scbasset} and three VAE-based models~\cite{xiong2019scale,xiong2022online,cui2024discrete}, which require self-supervised learning on each specific dataset. We conduct experiments on eight datasets from four tissue types and evaluate performance using four clustering metrics to ensure a fair and reliable comparison. 

All results presented in Table~\ref{clustering_table} reveal that ChromFound outperforms existing SOTA methods in all tissues and metrics, with average improvements of 17.02\%, 10.39\%, 6.72\% and 6.69\% in ARI, FMI, NMI and AMI, respectively. The greater improvements observed in ARI compared to NMI and AMI infer the robustness of ChromFound to technical noise such as low read depth and batch effects, as ARI is particularly sensitive to local structures within the cell representation space. We discuss ChromFound's ability to denoise low count and batch effect as below.

\begin{figure}[htbp]
  \includegraphics[width=\textwidth]{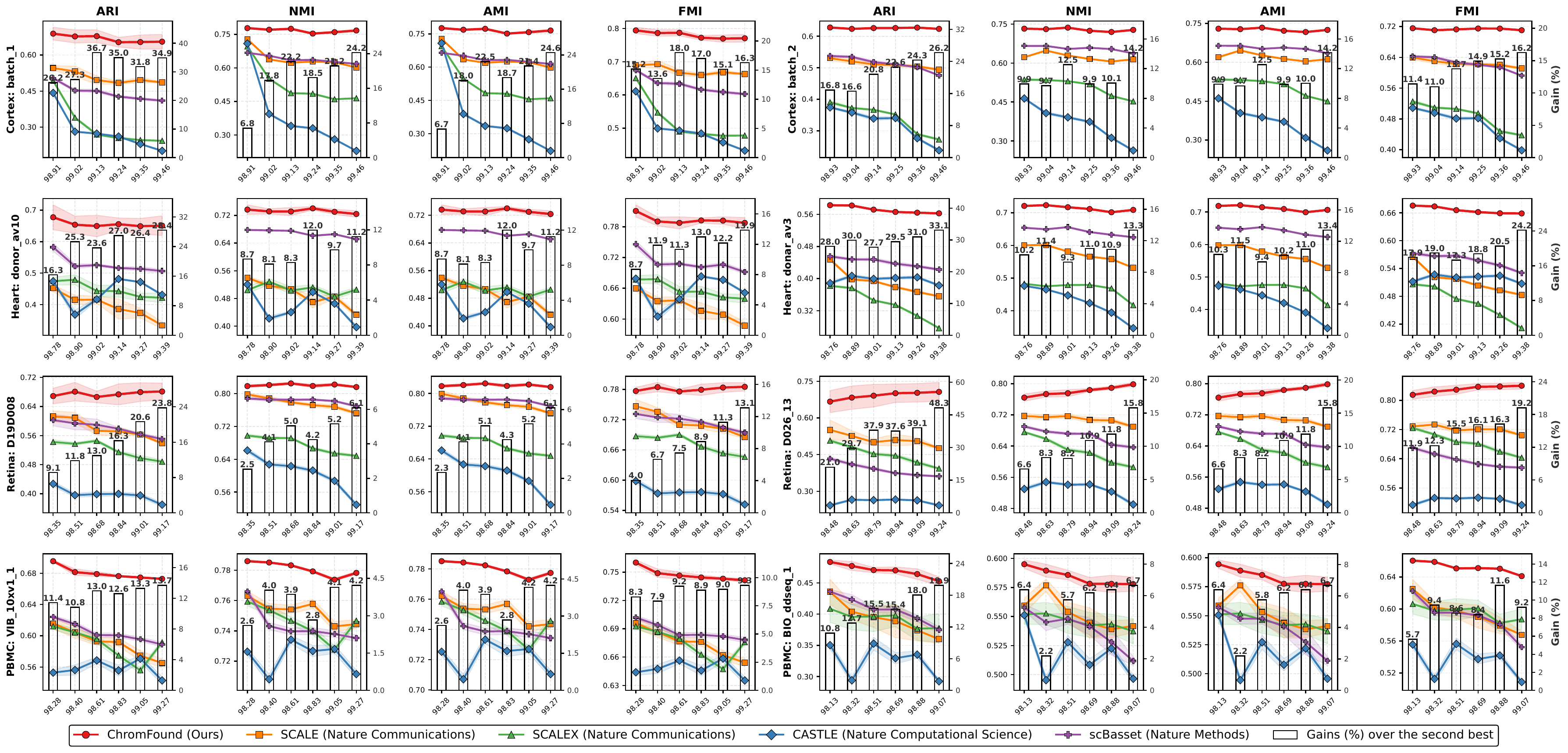}
  \caption{Results of denoising low count. The left y-axis shows the absolute metric values (scatter plot) and the right y-axis indicates the relative gains (\%) over the second-best methods (bar plot).}
  \label{dropout}
\end{figure}

\textbf{Denoising low count}: 
\label{denoise low count}
Some studies~\cite{chen2019assessment,li2021chromatin,tayyebi2024scalable} have acknowledged that the prevalence of missing signals in scATAC-seq makes standard analyses challenging. Although all datasets in Table \ref{clustering_table} already exhibit high sparsity (\textasciitilde99\%), we further simulate increasingly sparse conditions by downsampling accessible OCRs to 10\%, 20\%, 30\%, 40\% and 50\% of their original counts. We adopt the same benchmarking methods and clustering metrics as listed in Table~\ref{clustering_table}. As shown in Fig.~\ref{dropout}, ChromFound consistently maintains stable performance across all levels of downsampling, with its advantage over baseline methods becoming increasingly pronounced as the sparsity intensifies. Furthermore, we observe that ChromFound performs better on the retina dataset under stronger downsampling. This improvement likely reflects protocol-specific artifacts, as retina data contain more frequently accessible OCRs, introducing redundancy and noise. Random downsampling suppresses these excessive signals, enhances the signal-to-noise ratio, and yields the observed performance gain.

\textbf{Denoising batch effect}:
\label{denoise batch effect}
Denoising batch effects remains a major challenge in scATAC-seq analysis~\cite{miao2025depth,tayyebi2024scalable,scib}. We evaluate ChromFound on datasets from 4 tissues using biological conservation metrics~\cite{scib} consistent with scGPT~\cite{cui2024scgpt}. In addition to PeakVI~\cite{peakvi}, PoissonVI~\cite{poissonvi}, SCALEX~\cite{xiong2022online} and scBasset~\cite{yuan2022scbasset}, we include five specialized batch integration methods~\cite{scvi,scanorama,harmony,scanvi,liger} for comparison. As shown in Table~\ref{batch_effect_table}, ChromFound in zero-shot settings outperforms existing methods by more effectively mitigating batch effects while better preserving biologically significant variation.

\begin{table}[htbp]
\caption{Results of denoising batch effect. Result style: \textbf{best}, \uline{second best}, \textcolor{OliveGreen}{\textbf{relative gains}}. AvgBIO is the average of $\text{ARI}_{\text{cell}}$, $\text{NMI}_{\text{cell}}$ and $\text{ASW}_{\text{cell}}$ measuring biological consistency. AvgBATCH is computed as the average of $\text{ASW}_{\text{batch}}$ and GraphConn to summarize the batch mixing performance.}
\resizebox{\textwidth}{!}{%
\begin{tabular}{@{}c|cc|cc|cc|cc@{}}
\toprule
\multirow{2}{*}{\textbf{Model}} & \multicolumn{2}{c|}{\begin{tabular}[c]{@{}c@{}}\textbf{Bone} To326K~\cite{to2024multi}\\ batch 43/44\end{tabular}} & \multicolumn{2}{c|}{\begin{tabular}[c]{@{}c@{}}\textbf{Heart} Kuppe139K~\cite{kuppe2022spatial}\\ donar av3/av10\end{tabular}} & \multicolumn{2}{c|}{\begin{tabular}[c]{@{}c@{}}\textbf{PBMC} 169K~\cite{de2024systematic}\\ HAR ddseq 1/2\end{tabular}} & \multicolumn{2}{c}{\begin{tabular}[c]{@{}c@{}}\textbf{Cortex} Morabito130K~\cite{cortex63k}\\ batch 1/2\end{tabular}} \\
                       & AvgBIO($\uparrow$)                                     & AvgBATCH($\uparrow$)                                  & AvgBIO($\uparrow$)                                        & AvgBATCH($\uparrow$)                                      & AvgBIO($\uparrow$)                                     & AvgBATCH($\uparrow$)                                  & AvgBIO($\uparrow$)                                       & AvgBATCH($\uparrow$)                                     \\ \midrule
scVI~\cite{scvi}                   & 0.3713                                     & 0.9026                                    & 0.7303                                        & 0.8448                                        & 0.6101                                     & 0.7932                                    & \uline{0.7238}                                       & 0.9217                                       \\
PeakVI~\cite{peakvi}                   & 0.2981                                     & 0.9173                                    & 0.5657                                        & 0.7564                                        & 0.6172                                     & 0.7889                                    & 0.7256                                       & 0.9254                                       \\
PoissonVI~\cite{poissonvi}                   & 0.2936                                     & 0.9178                                    & 0.5621                                        & 0.7719                                        & 0.6029                                     & \uline{0.8081}                                    & 0.7240                                       & 0.9279                                       \\
Scanorama~\cite{scanorama}              & \uline{0.4386}                                     & 0.8722                                    & 0.5320                                        & 0.7779                                        & 0.6036                                     & 0.7964                                    & 0.6895                                       & 0.9243                                       \\

Harmony~\cite{harmony}                & 0.3789                                     & 0.8558                                    & 0.5119                                        & 0.8390                                        & \uline{0.6232}                                     & 0.7923                                    & 0.7152                                       & 0.9101                                       \\
scANVI~\cite{scanvi}                & 0.4351                                     & 0.8479                                    & 0.6810                                        & 0.8401                                        & 0.6028                                     & 0.8035                                    & 0.7185                                       & 0.9208                                       \\
Liger~\cite{liger}                & 0.1992                                     & 0.6471                                    & 0.7286                                        & 0.8019                                        & 0.6051                                     & 0.8009                                    & 0.6822                                       & 0.9208                                       \\
SCALEX~\cite{xiong2022online}                 & 0.1778                                     & 0.9177                                    & \uline{0.7596}                                        & 0.8415                                        & 0.5989                                     & 0.8004                                    & 0.4789                                       & 0.8728                                       \\
scBasset~\cite{yuan2022scbasset}               & 0.3650                                     & \uline{0.9208}                                    & 0.6216                                        & \uline{0.8591}                                        & 0.6024                                     & 0.8014                                    & 0.7045                                       & \uline{0.9301}                                       \\
\textbf{ChromFound}    & \textbf{0.6408}                            & \textbf{0.9289}                           & \textbf{0.8180}                               & \textbf{0.8679}                               & \textbf{0.6443}                            & \textbf{0.8217}                           & \textbf{0.7440}                              & \textbf{0.9565}                              \\ \midrule
\textbf{Gain(\%)}      & \textcolor{OliveGreen}{\textbf{46.07}}                             & \textcolor{OliveGreen}{\textbf{0.88}}                             & \textcolor{OliveGreen}{\textbf{7.69}}                                 & \textcolor{OliveGreen}{\textbf{1.03}}                                 & \textcolor{OliveGreen}{\textbf{3.37}}                              & \textcolor{OliveGreen}{\textbf{2.25}}                             & \textcolor{OliveGreen}{\textbf{2.78}}                                & \textcolor{OliveGreen}{\textbf{2.84}}                                \\ \bottomrule
\end{tabular}%
\label{batch_effect_table}
}
\end{table}

\subsection{Cell Type Annotation}
\label{sec:celltype_annotation}
\begin{figure}[htbp]
  \includegraphics[width=\textwidth]{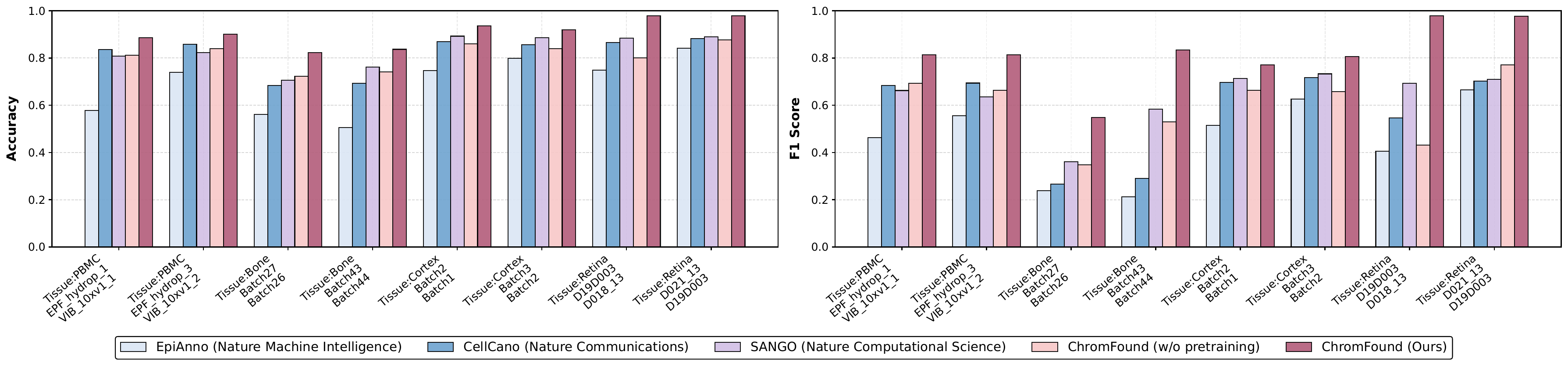}
  \caption{Results of cell type annotation, evaluated by Accuracy (left) and macro F1 Score (right).}
  \label{celltype_annotation}
\end{figure}

Computational cell type identification is a fundamental task in single-cell omics analysis. To adapt ChromFound for this task, we introduce a feature projection head followed by MLP layers as described in Appendix~\ref{cell_type_annotation_details}.  Our methods are compared to three strong baseline models designed for cell type annotation of scATAC-seq, Cellcano~\cite{ma2023cellcano}, EpiAnno~\cite{chen2022cell} and SANGO~\cite{zeng2024deciphering}. The variant of ChromFound trained from scratch is also included to assess the contribution of pretraining. Fig.~\ref{celltype_annotation} highlights that ChromFound consistently outperforms all benchmark methods, achieving the highest performance in macro F1 score and demonstrating its ability to accurately annotate diverse cell types. The confusion matrices on PBMC169K~\cite{de2024systematic} for ChromFound and CellCano~\cite{ma2023cellcano} are shown in Fig.~\ref{confusion_matrix}. The results indicate that Cellcano struggles to correctly classify rare cell types such as natural killer cells, CD16\textsuperscript{+} monocytes, and dendritic cells, likely due to their low abundance. In contrast, ChromFound significantly improves the classification accuracy for these cell types, enabling a more precise understanding of immune cell composition and dynamics.

\subsection{Cross-omics Prediction}
\label{sec:modality_prediction}
Although recent advances have enabled the simultaneous profiling of multiple assays, most single-cell datasets remain isolated, posing significant challenges for effective multi-omics analysis. To explore the potential of ChromFound in multi-omics prediction, we utilize scATAC-seq data as the source modality and scRNA-seq data as the target modality. Implementation details are provided in Appendix~\ref{modality_prediction_appendix}. For comprehensive evaluation, we benchmark on five paired scATAC-seq and scRNA-seq datasets~\cite{zhu2023multi,to2024multi,luecken2021sandbox,open-problems-multimodal,zhu2023single} using two standard correlation metrics. Three representative baseline methods~\cite{wu2021babel,yang2021multi,wen2022graph} are included with default configurations from the DANCE toolkit~\cite{ding2022dance}. As shown in Table~\ref{cross_omics}, ChromFound consistently outperforms all baselines across datasets in both PCC and CCC, highlighting its strong capability in cross-omics prediction. These results highlight the potential of ChromFound to facilitate biological analyses such as inferring gene-enhancer relationships and predicting transcriptional response to enhancer perturbation. The two biological applications are discussed in detail below.

\begin{table}[htbp]
\caption{Cross-omics prediction results. Result style: \textbf{best}, \uline{second best}, \textcolor{OliveGreen}{\textbf{relative gains}}. ChromFound* stands for the version of ChromFound without loading the pretrained checkpoint.}
\renewcommand{\arraystretch}{0.8}
\scriptsize
\resizebox{\textwidth}{!}{%
\begin{tabular}{@{}c|cc|cc|cc|cc|cc@{}}
\toprule
\multirow{3}{*}{\textbf{Model}} & \multicolumn{2}{c|}{\multirow{2}{*}{\begin{tabular}[c]{@{}c@{}}\textbf{Cortex}\\ Zhu45K~\cite{zhu2023multi}\end{tabular}}} & \multicolumn{2}{c|}{\multirow{2}{*}{\begin{tabular}[c]{@{}c@{}}\textbf{Bone}\\ To326K~\cite{to2024multi}\end{tabular}}} & \multicolumn{2}{c|}{\multirow{2}{*}{\begin{tabular}[c]{@{}c@{}}\textbf{BMMC}\\ multiome 2021~\cite{luecken2021sandbox}\end{tabular}}} & \multicolumn{2}{c|}{\multirow{2}{*}{\begin{tabular}[c]{@{}c@{}}\textbf{BMMC}\\ atac2gex 2022~\cite{open-problems-multimodal}\end{tabular}}} & \multicolumn{2}{c}{\multirow{2}{*}{\begin{tabular}[c]{@{}c@{}}\textbf{Cell lines}\\ Zhu11K~\cite{zhu2023single}\end{tabular}}} \\
                                & \multicolumn{2}{c|}{}                                                                                  & \multicolumn{2}{c|}{}                                                                                & \multicolumn{2}{c|}{}                                                                                       & \multicolumn{2}{c|}{}                                                                                       & \multicolumn{2}{c}{}                                                                               \\
                                & PCC($\uparrow$)                                                & CCC($\uparrow$)                                                & PCC($\uparrow$)                                                & CCC($\uparrow$)                                               & PCC($\uparrow$)                                                   & CCC($\uparrow$)                                                   & PCC$\uparrow$                                                   & CCC($\uparrow$)                                                   & PCC($\uparrow$)                                               & CCC($\uparrow$)                                              \\ \midrule
BABEL~\cite{wu2021babel}                          & 0.7975                                             & \uline{0.7687}                                            & \uline{0.8020}                                            & 0.7223                                           & 0.3854                                               & 0.3695                                               & 0.8901                                               & 0.8329                                               & \uline{0.9196}                                           & \uline{0.8608}                                          \\
CMAE~\cite{yang2021multi}                           & 0.7973                                             & 0.7525                                            & 0.7986                                            & 0.6683                                           & 0.3435                                               & 0.3204                                               & 0.8983                                               & 0.7990                                               & 0.9136                                           & 0.8539                                          \\
scMM~\cite{minoura2021scmm}                           & 0.7793                                             & 0.7325                                            & 0.7586                                            & 0.6513                                           & 0.3718                                               & 0.2983                                               & 0.8321                                               & 0.7027                                               & 0.8587                                           & 0.8062                                          \\
scMoGNN~\cite{wen2022graph}                        & \uline{0.8001}                                             & 0.7210                                            & 0.7957                                            & \uline{0.7233}                                           & \uline{0.4168}                                               & \uline{0.3911}                                               & \uline{0.9124}                                               & \uline{0.8655}                                               & 0.7976                                           & 0.7263                                          \\ \midrule
ChromFound*         & 0.7864                                             & 0.7674                                            & 0.7887                                            & 0.6919                                           & 0.4213                                               & 0.3989                                               & 0.9279                                               & 0.8704                                               & 0.8992                                           & 0.8501                                          \\
\textbf{ChromFound}             & \textbf{0.8316}                                    & \textbf{0.8064}                                   & \textbf{0.8304}                                   & \textbf{0.7472}                                  & \textbf{0.4249}                                      & \textbf{0.4032}                                      & \textbf{0.9293}                                      & \textbf{0.8818}                                      & \textbf{0.9449}                                  & \textbf{0.9071}                                 \\ \midrule
\textbf{Gain(\%)}                        & \textcolor{OliveGreen}{\textbf{4.10}}                                               & \textcolor{OliveGreen}{\textbf{4.67}}                                              & \textcolor{OliveGreen}{\textbf{3.54}}                                              & \textcolor{OliveGreen}{\textbf{3.30}}                                             & \textcolor{OliveGreen}{\textbf{1.94}}                                                 & \textcolor{OliveGreen}{\textbf{3.09}}                                                 & \textcolor{OliveGreen}{\textbf{1.85}}                                                 & \textcolor{OliveGreen}{\textbf{1.88}}                                                 & \textcolor{OliveGreen}{\textbf{2.67}}                                             & \textcolor{OliveGreen}{\textbf{5.11}}                                            \\ \bottomrule
\end{tabular}%
\label{cross_omics}
}
\end{table}

\textbf{Biological Application: Predicting enhancer-gene link and perturbation response}
\label{biological application}

\begin{figure}[htbp]
    \subfigure[]{
    \label{roc_curves}
    \includegraphics[width =0.48\textwidth]{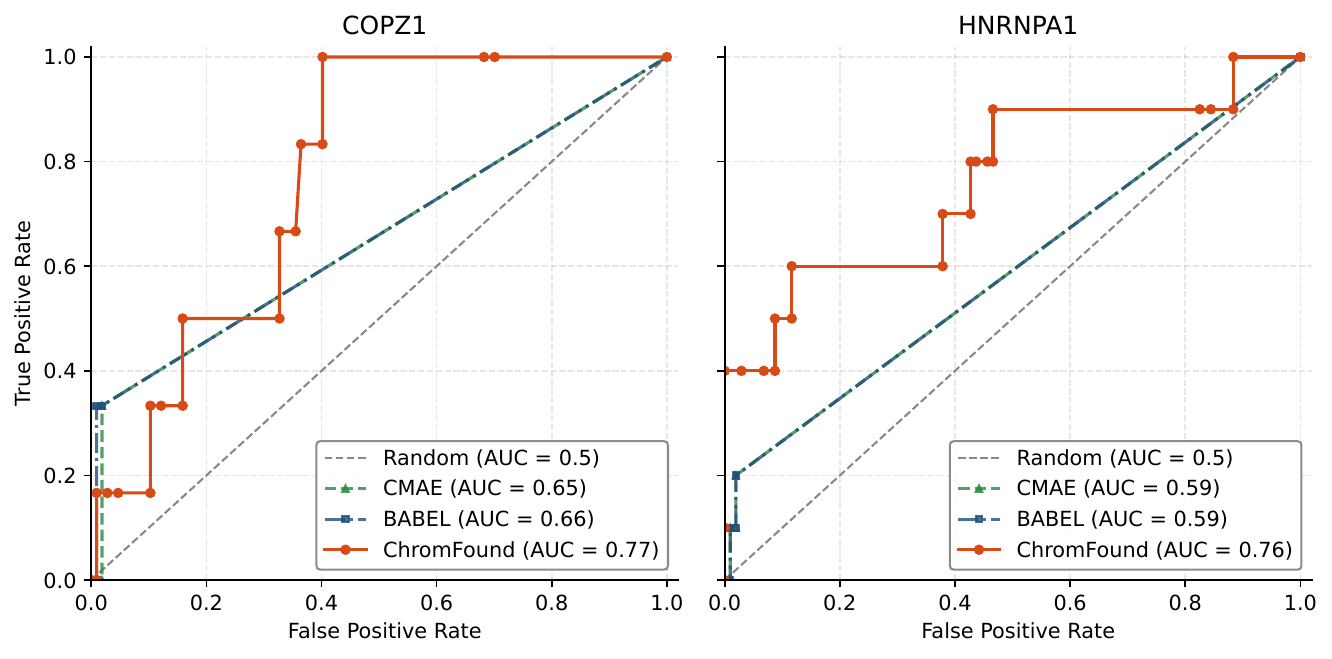}
    }
    \subfigure[]
    {
    \label{perturbation}
    \includegraphics[width =0.48\textwidth]{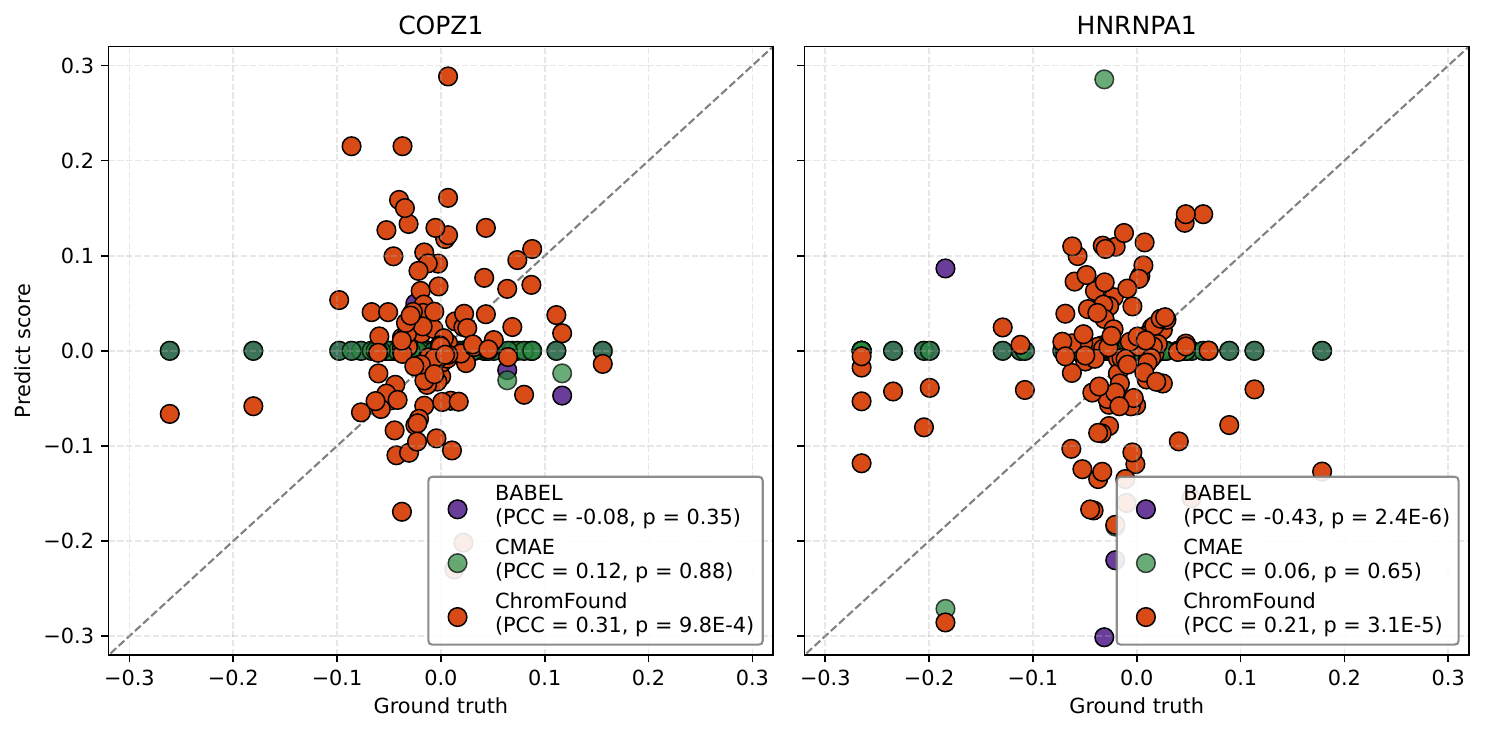}
    }
    \caption{\textbf{(a)}~ROC curves illustrate the performance of enhancer-gene link prediction, where true positive rates are plotted against false positive rates across varying prediction thresholds. Ground truth labels are defined based on the significance (FDR < 0.05) of expression changes after enhancer perturbation. \textbf{(b)}~Scatter plots represent a specific enhancer perturbation. The y-axis shows the predicted scores of average gene responses of enhancer perturbation, while the x-axis represents the real effects of post-perturbation. Both magnitudes are rescaled to the unit norm.}
    \label{bio_figure}
\end{figure}

Enhancer elements in the human genome harbor thousands of genetic variants that regulate gene expression and contribute to the risk of common diseases~\cite{script}.
In our approach detailed in Appendix~\ref{biological_app_appendix}, we simulate OCR knockdown and estimate the resulting changes in gene expression. The absolute change reflects the likelihood of an enhancer-gene regulatory link, while the relative change captures the direction and strength of the transcriptional response.
We perform experiments on 141 K562 cells from the test split in Zhu11K~\cite{zhu2023single}, using the cross-omics model trained as evaluated in Table~\ref{cross_omics}. 
Fulco4K~\cite{zhu2023single} provides the ground truth of enhancer-gene links and quantitative effects through CRISPRi perturbations in K562 human erythroleukemia cells.
For clarity, the evaluation focuses on the COPZ1 and HNRNPA1 genes, which are involved in cancer-related processes such as stress adaptation and RNA splicing regulation~\cite{shtutman2011tumor, clower2010alternative}. BABEL~\cite{wu2021babel} and CMAE~\cite{yang2021multi} are included as benchmark methods due to their strong performance in cross-omics prediction in Zhu11K~\cite{zhu2023single}. 
As shown in Fig.~\ref{roc_curves}, ChromFound achieves the highest AUC scores with 0.77 for COPZ1 and 0.61 for HNRNPA1, accurately identifying enhancer-gene links. 
Fig.~\ref{perturbation} further shows that ChromFound better predicts transcriptional responses to enhancer perturbations. In contrast, BABEL~\cite{wu2021babel} and CMAE~\cite{yang2021multi} filter for the top 10,000 highly variable OCRs~\cite{ding2022dance}, leading to zero predictions of most enhancer perturbations. This limitation is evident in both Fig.\ref{roc_curves} and Fig.\ref{perturbation}, highlighting the advantage of ChromFound’s genome-wide modeling to support a more comprehensive understanding of regulatory mechanisms.

\subsection{Ablation study}
\label{sec:ablation_study}
\definecolor{mamba}{RGB}{230,245,255}   
\definecolor{modinnov}{RGB}{229,245,224} 
\definecolor{datascale}{RGB}{255,247,200}  
\definecolor{fullmodel}{RGB}{254,235,226}      

In this section, we conduct an ablation study to address three questions using the PBMC169K~\cite{de2024systematic}, sampled from batch VIB\_10xv1\_1. The results of cell representation are summarized in Table~\ref{ablation_study_table}.

\colorbox{modinnov}{Question 1}: Does our proposed ChromFound architecture contribute to performance?

We evaluate the individual impact of each key module in our model, including genome-aware tokenization (Section~\ref{tokenization}), the WPSA layer (Section~\ref{sec:wpsa}) and the Mamba block (Section~\ref{sec:mamba}). In rows 2-4 of Table~\ref{ablation_study_table}, we observe a clear performance drop in all clustering metrics compared to row 1, demonstrating that each component contributes complementarily to overall performance.

\colorbox{datascale}{Question 2}: Does scaling to large-scale data provide performance gains?

As shown in rows 5-6 of Table~\ref{ablation_study_table}, reducing the data size from 1.97 million to 0.2 million and 20 thousand cells leads to consistent degradation in all evaluation metrics, confirming that large-scale data learning facilitates better generalization.

\colorbox{mamba}{Question 3}: Is long-context modeling necessary for high-quality representations?

As seen in rows 7-8 of Table~\ref{ablation_study_table}, reducing the input length by half and by a quarter significantly hurts performance, highlighting the importance of long-context modeling for dynamic OCR landscapes.

\begin{table}[htbp]
\caption{Ablation study. \colorbox{fullmodel}{Red row} indicates the full model of ChromFound. The other colored rows correspond to the topics discussed in Section \ref{sec:ablation_study}.}
\renewcommand{\arraystretch}{1.0}
\resizebox{\textwidth}{!}{%
\begin{tabular}{@{}cccccc|cccc@{}}
\toprule
\textbf{\begin{tabular}[c]{@{}c@{}}Genome\\ info\end{tabular}} & \textbf{WPSA} & \textbf{Mamba} & \textbf{\begin{tabular}[c]{@{}c@{}}Data size\\ (million)\end{tabular}} & \textbf{\begin{tabular}[c]{@{}c@{}}\# of OCRs \\ per cell\end{tabular}} & \textbf{\begin{tabular}[c]{@{}c@{}}FLOPs(10e9)\\ per cell\end{tabular}} & \textbf{ARI}($\uparrow$)  & \textbf{FMI}($\uparrow$)  & \textbf{NMI}($\uparrow$)  & \textbf{AMI}($\uparrow$)  \\ \midrule
\rowcolor{fullmodel}
\ding{52}                                                                       & \ding{52}             & \ding{52}              & 1.97                                                                   & 440000                                                                    & 19.72                                                                     & \textbf{0.6953±0.0035} & \textbf{0.7601±0.0023} & \textbf{0.7860±0.0007} & \textbf{0.7852±0.0008} \\ \midrule
\rowcolor{modinnov}                                                                       & \ding{52}             & \ding{52}              & 1.97                                                  & 440000                                                   & 19.72                                                    & 0.6452±0.0149 & 0.7095±0.0098 & 0.7227±0.0021 & 0.7301±0.0017 \\
\rowcolor{modinnov}
                                                                       &              & \ding{52}              & 1.97                                                                       & 440000                                                                          & 19.72                                                                          & 0.5897±0.0086 & 0.6401±0.0113 & 0.6419±0.0012 & 0.6476±0.0015 \\
\rowcolor{modinnov}                                                                       & \ding{52}             &               & 1.97                                                                       & 440000                                                                          & 19.72                                                                          & 0.3075±0.0019 & 0.3098±0.0011 & 0.3101±0.0011 & 0.3326±0.0013 \\ \midrule
\rowcolor{datascale}
\ding{52}                                                                       & \ding{52}             & \ding{52}              & 0.2                                                                    & 440000                                                   & 19.72                                                    & 0.6539±0.0057 & 0.7261±0.0037 & 0.7593±0.0012 & 0.7583±0.0012 \\
\rowcolor{datascale}
\ding{52}                                                                       & \ding{52}             & \ding{52}              & 0.02                                                                   & 440000                                                                          & 19.72                                                                          & 0.6142±0.0053 & 0.6995±0.0034 & 0.7354±0.0015 & 0.7345±0.0015 \\ \midrule
\rowcolor{mamba}
\ding{52}                                                                       & \ding{52}             & \ding{52}              & 1.97                                                  & 220000                                                                    & 9.87                                                                      & 0.4012±0.0047 & 0.4500±0.0031 & 0.4535±0.0007 & 0.4425±0.0007 \\
\rowcolor{mamba}
\ding{52}                                                                       & \ding{52}             & \ding{52}              & 1.97                                                                       & 110000                                                                    & 4.93                                                                      & 0.3276±0.0051 & 0.3589±0.0048 & 0.3674±0.0002 & 0.3668±0.0002 \\ \bottomrule
\end{tabular}%
}
\label{ablation_study_table}
\end{table}

\section{Conclusion}
\label{conclusion}
In this work, we present ChromFound, to the best of our knowledge, the first foundation model specifically for scATAC-seq data. To address the inherent challenges of high-dimensionality, sparsity, and dynamic chromatin landscapes, we utilize a hybrid architecture that integrates short- and long-range dependencies with a genome-aware tokenization. Trained on 1.97 million single-cell profiles comprising over 30 human tissue types and 6 disease categories, ChromFound achieves SOTA performance in 6 downstream tasks, showing great robustness in zero-shot cell representation. Its biological significance is further demonstrated by its ability to accurately infer transcriptional responses to enhancer perturbations.
In the future, we aim to explore the broader applicability of ChromFound in the large-scale mapping of enhancer-gene regulatory interactions, a critical yet largely uncharted component of the \textit{cis}-regulatory landscape of the human genome.

\section{Acknowledgments}
The authors gratefully acknowledge the following funding sources: AI for Science Program, Shanghai Municipal Commission of Economy and Information, National Natural Science Foundation of China (Grant No. 82394432, 92249302), Shanghai Municipal Science and Technology Major Project (2023SHZDZX02), the King Abdullah University of Science and Technology (KAUST) Office of Research Administration (ORA) under Award No REI/1/5234-01-01, REI/1/5414-01-01, REI/1/5289-01-01, REI/1/5404-01-01, REI/1/5992-01-01, URF/1/4663-01-01, Center of Excellence for Smart Health (KCSH) under award number 5932, and Center of Excellence on Generative AI under award number 5940. The computations in this research were performed using the Computing for the Future at Fudan (CFFF) platform of Fudan University.

{\small
\bibliographystyle{ieee_fullname}
\bibliography{citation}

\begin{thebibliography}{10}\itemsep=-1pt

\bibitem{peakvi}
Tal Ashuach, Daniel~A Reidenbach, Adam Gayoso, and Nir Yosef.
\newblock Peakvi: A deep generative model for single-cell chromatin accessibility analysis.
\newblock {\em Cell reports methods}, 2(3), 2022.

\bibitem{Bonev2016OrganizationAF}
Boyan~B. Bonev and Giacomo Cavalli.
\newblock Organization and function of the 3d genome.
\newblock {\em Nature Reviews Genetics}, 17:661--678, 2016.

\bibitem{buenrostro3k}
Jason~D Buenrostro, M~Ryan Corces, Caleb~A Lareau, Beijing Wu, Alicia~N Schep, Martin~J Aryee, Ravindra Majeti, Howard~Y Chang, and William~J Greenleaf.
\newblock Integrated single-cell analysis maps the continuous regulatory landscape of human hematopoietic differentiation.
\newblock {\em Cell}, 173(6):1535--1548, 2018.

\bibitem{buenrostro2015single}
Jason~D Buenrostro, Beijing Wu, Ulrike~M Litzenburger, Dave Ruff, Michael~L Gonzales, Michael~P Snyder, Howard~Y Chang, and William~J Greenleaf.
\newblock Single-cell chromatin accessibility reveals principles of regulatory variation.
\newblock {\em Nature}, 523(7561):486--490, 2015.

\bibitem{open-problems-multimodal}
Daniel Burkhardt, Malte Luecken, Andrew Benz, Peter Holderrieth, Jonathan Bloom, Christopher Lance, Ashley Chow, and Ryan Holbrook.
\newblock Open problems - multimodal single-cell integration.
\newblock \url{https://kaggle.com/competitions/open-problems-multimodal}, 2022.
\newblock Kaggle.

\bibitem{chen2019assessment}
Huidong Chen, Caleb Lareau, Tommaso Andreani, Michael~E Vinyard, Sara~P Garcia, Kendell Clement, Miguel~A Andrade-Navarro, Jason~D Buenrostro, and Luca Pinello.
\newblock Assessment of computational methods for the analysis of single-cell atac-seq data.
\newblock {\em Genome biology}, 20:1--25, 2019.

\bibitem{chen2022cell}
Xiaoyang Chen, Shengquan Chen, Shuang Song, Zijing Gao, Lin Hou, Xuegong Zhang, Hairong Lv, and Rui Jiang.
\newblock Cell type annotation of single-cell chromatin accessibility data via supervised bayesian embedding.
\newblock {\em Nature Machine Intelligence}, 4(2):116--126, 2022.

\bibitem{clower2010alternative}
Cynthia~V Clower, Deblina Chatterjee, Zhenxun Wang, Lewis~C Cantley, Matthew~G Vander~Heiden, and Adrian~R Krainer.
\newblock The alternative splicing repressors hnrnp a1/a2 and ptb influence pyruvate kinase isoform expression and cell metabolism.
\newblock {\em Proceedings of the National Academy of Sciences}, 107(5):1894--1899, 2010.

\bibitem{corces0.5k}
M~Ryan Corces, Jason~D Buenrostro, Beijing Wu, Peyton~G Greenside, Steven~M Chan, Julie~L Koenig, Michael~P Snyder, Jonathan~K Pritchard, Anshul Kundaje, William~J Greenleaf, et~al.
\newblock Lineage-specific and single-cell chromatin accessibility charts human hematopoiesis and leukemia evolution.
\newblock {\em Nature genetics}, 48(10):1193--1203, 2016.

\bibitem{corces2020single}
M~Ryan Corces, Anna Shcherbina, Soumya Kundu, Michael~J Gloudemans, Laure Fr{\'e}sard, Jeffrey~M Granja, Bryan~H Louie, Tiffany Eulalio, Shadi Shams, S~Tansu Bagdatli, et~al.
\newblock Single-cell epigenomic analyses implicate candidate causal variants at inherited risk loci for alzheimer’s and parkinson’s diseases.
\newblock {\em Nature genetics}, 52(11):1158--1168, 2020.

\bibitem{cui2024scgpt}
Haotian Cui, Chloe Wang, Hassaan Maan, Kuan Pang, Fengning Luo, Nan Duan, and Bo Wang.
\newblock scgpt: toward building a foundation model for single-cell multi-omics using generative ai.
\newblock {\em Nature Methods}, pages 1--11, 2024.

\bibitem{cui2024discrete}
Xuejian Cui, Xiaoyang Chen, Zhen Li, Zijing Gao, Shengquan Chen, and Rui Jiang.
\newblock Discrete latent embedding of single-cell chromatin accessibility sequencing data for uncovering cell heterogeneity.
\newblock {\em Nature Computational Science}, pages 1--14, 2024.

\bibitem{cusanovich2015multiplex}
Darren~A Cusanovich, Riza Daza, Andrew Adey, Hannah~A Pliner, Lena Christiansen, Kevin~L Gunderson, Frank~J Steemers, Cole Trapnell, and Jay Shendure.
\newblock Multiplex single-cell profiling of chromatin accessibility by combinatorial cellular indexing.
\newblock {\em Science}, 348(6237):910--914, 2015.

\bibitem{cusanovich2018cis}
Darren~A Cusanovich, James~P Reddington, David~A Garfield, Riza~M Daza, Delasa Aghamirzaie, Raquel Marco-Ferreres, Hannah~A Pliner, Lena Christiansen, Xiaojie Qiu, Frank~J Steemers, et~al.
\newblock The cis-regulatory dynamics of embryonic development at single-cell resolution.
\newblock {\em Nature}, 555(7697):538--542, 2018.

\bibitem{de2024systematic}
Florian~V De~Rop, Gert Hulselmans, Chris Flerin, Paula Soler-Vila, Albert Rafels, Valerie Christiaens, Carmen~Bravo Gonz{\'a}lez-Blas, Domenica Marchese, Ginevra Caratu, Suresh Poovathingal, et~al.
\newblock Systematic benchmarking of single-cell atac-sequencing protocols.
\newblock {\em Nature biotechnology}, 42(6):916--926, 2024.

\bibitem{ding2022dance}
Jiayuan Ding, Hongzhi Wen, Wenzhuo Tang, Renming Liu, Zhaoheng Li, Julian Venegas, Runze Su, Dylan Molho, Wei Jin, Wangyang Zuo, et~al.
\newblock Dance: A deep learning library and benchmark for single-cell analysis.
\newblock {\em bioRxiv}, pages 2022--10, 2022.

\bibitem{domcke2020human}
Silvia Domcke, Andrew~J Hill, Riza~M Daza, Junyue Cao, Diana~R O’Day, Hannah~A Pliner, Kimberly~A Aldinger, Dmitry Pokholok, Fan Zhang, Jennifer~H Milbank, et~al.
\newblock A human cell atlas of fetal chromatin accessibility.
\newblock {\em Science}, 370(6518):eaba7612, 2020.

\bibitem{fu2025foundation}
Xi Fu, Shentong Mo, Alejandro Buendia, Anouchka~P Laurent, Anqi Shao, Maria del~Mar Alvarez-Torres, Tianji Yu, Jimin Tan, Jiayu Su, Romella Sagatelian, et~al.
\newblock A foundation model of transcription across human cell types.
\newblock {\em Nature}, pages 1--9, 2025.

\bibitem{fulco2019activity}
Charles~P Fulco, Joseph Nasser, Thouis~R Jones, Glen Munson, Drew~T Bergman, Vidya Subramanian, Sharon~R Grossman, Rockwell Anyoha, Benjamin~R Doughty, Tejal~A Patwardhan, et~al.
\newblock Activity-by-contact model of enhancer--promoter regulation from thousands of crispr perturbations.
\newblock {\em Nature genetics}, 51(12):1664--1669, 2019.

\bibitem{ghazanfar2024stabilized}
Shila Ghazanfar, Carolina Guibentif, and John~C Marioni.
\newblock Stabilized mosaic single-cell data integration using unshared features.
\newblock {\em Nature biotechnology}, 42(2):284--292, 2024.

\bibitem{gong2024xtrimogene}
Jing Gong, Minsheng Hao, Xingyi Cheng, Xin Zeng, Chiming Liu, Jianzhu Ma, Xuegong Zhang, Taifeng Wang, and Le Song.
\newblock xtrimogene: an efficient and scalable representation learner for single-cell rna-seq data.
\newblock {\em Advances in Neural Information Processing Systems}, 36, 2024.

\bibitem{granja2021archr}
Jeffrey~M Granja, M~Ryan Corces, Sarah~E Pierce, S~Tansu Bagdatli, Hani Choudhry, Howard~Y Chang, and William~J Greenleaf.
\newblock Archr is a scalable software package for integrative single-cell chromatin accessibility analysis.
\newblock {\em Nature genetics}, 53(3):403--411, 2021.

\bibitem{pbmc10k}
Jeffrey~M Granja, Sandy Klemm, Lisa~M McGinnis, Arwa~S Kathiria, Anja Mezger, M~Ryan Corces, Benjamin Parks, Eric Gars, Michaela Liedtke, Grace~XY Zheng, et~al.
\newblock Single-cell multiomic analysis identifies regulatory programs in mixed-phenotype acute leukemia.
\newblock {\em Nature biotechnology}, 37(12):1458--1465, 2019.

\bibitem{gu2023mamba}
Albert Gu and Tri Dao.
\newblock Mamba: Linear-time sequence modeling with selective state spaces.
\newblock {\em arXiv preprint arXiv:2312.00752}, 2023.

\bibitem{gu2021efficiently}
Albert Gu, Karan Goel, and Christopher R{\'e}.
\newblock Efficiently modeling long sequences with structured state spaces.
\newblock {\em arXiv preprint arXiv:2111.00396}, 2021.

\bibitem{gu2021combining}
Albert Gu, Isys Johnson, Karan Goel, Khaled Saab, Tri Dao, Atri Rudra, and Christopher R{\'e}.
\newblock Combining recurrent, convolutional, and continuous-time models with linear state space layers.
\newblock {\em Advances in neural information processing systems}, 34:572--585, 2021.

\bibitem{hao2024large}
Minsheng Hao, Jing Gong, Xin Zeng, Chiming Liu, Yucheng Guo, Xingyi Cheng, Taifeng Wang, Jianzhu Ma, Xuegong Zhang, and Le Song.
\newblock Large-scale foundation model on single-cell transcriptomics.
\newblock {\em Nature Methods}, pages 1--11, 2024.

\bibitem{hao2021integrated}
Yuhan Hao, Stephanie Hao, Erica Andersen-Nissen, William~M Mauck, Shiwei Zheng, Andrew Butler, Maddie~J Lee, Aaron~J Wilk, Charlotte Darby, Michael Zager, et~al.
\newblock Integrated analysis of multimodal single-cell data.
\newblock {\em Cell}, 184(13):3573--3587, 2021.

\bibitem{scanorama}
Brian Hie, Bryan Bryson, and Bonnie Berger.
\newblock Efficient integration of heterogeneous single-cell transcriptomes using scanorama.
\newblock {\em Nature biotechnology}, 37(6):685--691, 2019.

\bibitem{kim2015architectural}
Tae-Kyung Kim and Ramin Shiekhattar.
\newblock Architectural and functional commonalities between enhancers and promoters.
\newblock {\em Cell}, 162(5):948--959, 2015.

\bibitem{Klemm2019ChromatinAA}
Sandy~L. Klemm, Zohar Shipony, and William~James Greenleaf.
\newblock Chromatin accessibility and the regulatory epigenome.
\newblock {\em Nature Reviews Genetics}, 20:207 -- 220, 2019.

\bibitem{harmony}
Ilya Korsunsky, Nghia Millard, Jean Fan, Kamil Slowikowski, Fan Zhang, Kevin Wei, Yuriy Baglaenko, Michael Brenner, Po-ru Loh, and Soumya Raychaudhuri.
\newblock Fast, sensitive and accurate integration of single-cell data with harmony.
\newblock {\em Nature methods}, 16(12):1289--1296, 2019.

\bibitem{kuppe2022spatial}
Christoph Kuppe, Ricardo~O Ramirez~Flores, Zhijian Li, Sikander Hayat, Rebecca~T Levinson, Xian Liao, Monica~T Hannani, Jovan Tanevski, Florian W{\"u}nnemann, James~S Nagai, et~al.
\newblock Spatial multi-omic map of human myocardial infarction.
\newblock {\em Nature}, 608(7924):766--777, 2022.

\bibitem{li2021chromatin}
Zhijian Li, Christoph Kuppe, Susanne Ziegler, Mingbo Cheng, Nazanin Kabgani, Sylvia Menzel, Martin Zenke, Rafael Kramann, and Ivan~G Costa.
\newblock Chromatin-accessibility estimation from single-cell atac-seq data with scopen.
\newblock {\em Nature communications}, 12(1):6386, 2021.

\bibitem{liang2023multi}
Qingnan Liang, Xuesen Cheng, Jun Wang, Leah Owen, Akbar Shakoor, John~L Lillvis, Charles Zhang, Michael Farkas, Ivana~K Kim, Yumei Li, et~al.
\newblock A multi-omics atlas of the human retina at single-cell resolution.
\newblock {\em Cell genomics}, 3(6), 2023.

\bibitem{liu2024mamba4rec}
Chengkai Liu, Jianghao Lin, Jianling Wang, Hanzhou Liu, and James Caverlee.
\newblock Mamba4rec: Towards efficient sequential recommendation with selective state space models.
\newblock {\em arXiv preprint arXiv:2403.03900}, 2024.

\bibitem{liger}
Jialin Liu, Chao Gao, Joshua Sodicoff, Velina Kozareva, Evan~Z Macosko, and Joshua~D Welch.
\newblock Jointly defining cell types from multiple single-cell datasets using liger.
\newblock {\em Nature protocols}, 15(11):3632--3662, 2020.

\bibitem{scvi}
Romain Lopez, Jeffrey Regier, Michael~B Cole, Michael~I Jordan, and Nir Yosef.
\newblock Deep generative modeling for single-cell transcriptomics.
\newblock {\em Nature methods}, 15(12):1053--1058, 2018.

\bibitem{loshchilov2017decoupled}
I Loshchilov.
\newblock Decoupled weight decay regularization.
\newblock {\em arXiv preprint arXiv:1711.05101}, 2017.

\bibitem{luecken2021sandbox}
Malte~D Luecken, Daniel~Bernard Burkhardt, Robrecht Cannoodt, Christopher Lance, Aditi Agrawal, Hananeh Aliee, Ann~T Chen, Louise Deconinck, Angela~M Detweiler, Alejandro~A Granados, et~al.
\newblock A sandbox for prediction and integration of dna, rna, and proteins in single cells.
\newblock In {\em Thirty-fifth conference on neural information processing systems datasets and benchmarks track (Round 2)}, 2021.

\bibitem{scib}
Malte~D Luecken, Maren B{\"u}ttner, Kridsadakorn Chaichoompu, Anna Danese, Marta Interlandi, Michaela~F M{\"u}ller, Daniel~C Strobl, Luke Zappia, Martin Dugas, Maria Colom{\'e}-Tatch{\'e}, et~al.
\newblock Benchmarking atlas-level data integration in single-cell genomics.
\newblock {\em Nature methods}, 19(1):41--50, 2022.

\bibitem{ma2023cellcano}
Wenjing Ma, Jiaying Lu, and Hao Wu.
\newblock Cellcano: supervised cell type identification for single cell atac-seq data.
\newblock {\em Nature Communications}, 14(1):1864, 2023.

\bibitem{martens2024modeling}
Laura~D Martens, David~S Fischer, Vicente~A Y{\'e}pez, Fabian~J Theis, and Julien Gagneur.
\newblock Modeling fragment counts improves single-cell atac-seq analysis.
\newblock {\em Nature Methods}, 21(1):28--31, 2024.

\bibitem{poissonvi}
Laura~D Martens, David~S Fischer, Vicente~A Y{\'e}pez, Fabian~J Theis, and Julien Gagneur.
\newblock Modeling fragment counts improves single-cell atac-seq analysis.
\newblock {\em Nature Methods}, 21(1):28--31, 2024.

\bibitem{cpeaks}
Qiuchen Meng, Xinze Wu, Wenchang Chen, Yubo Zhao, Chen Li, Zheng Wei, Jiaqi Li, Xi Xi, Sijie Chen, Catherine Zhang, et~al.
\newblock A generic reference defined by consensus peaks for scatac-seq data analysis.
\newblock {\em bioRxiv}, pages 2023--05, 2023.

\bibitem{miao2025depth}
Zhen Miao, Jianqiao Wang, Kernyu Park, Da Kuang, and Junhyong Kim.
\newblock Depth-corrected multi-factor dissection of chromatin accessibility for scatac-seq data with pacs.
\newblock {\em Nature Communications}, 16(1):1--15, 2025.

\bibitem{minoura2021scmm}
Kodai Minoura, Ko Abe, Hyunha Nam, Hiroyoshi Nishikawa, and Teppei Shimamura.
\newblock scmm: Mixture-of-experts multimodal deep generative model for single-cell multiomics data analysis.
\newblock {\em bioRxiv}, pages 2021--02, 2021.

\bibitem{cortex63k}
Samuel Morabito, Emily Miyoshi, Neethu Michael, Saba Shahin, Alessandra~Cadete Martini, Elizabeth Head, Justine Silva, Kelsey Leavy, Mari Perez-Rosendahl, and Vivek Swarup.
\newblock Single-nucleus chromatin accessibility and transcriptomic characterization of alzheimer’s disease.
\newblock {\em Nature genetics}, 53(8):1143--1155, 2021.

\bibitem{cortex2k}
Ryan~M Mulqueen, Dmitry Pokholok, Brendan~L O’Connell, Casey~A Thornton, Fan Zhang, Brian~J O’Roak, Jason Link, Galip~G{\"u}rkan Yard{\i}mc{\i}, Rosalie~C Sears, Frank~J Steemers, et~al.
\newblock High-content single-cell combinatorial indexing.
\newblock {\em Nature biotechnology}, 39(12):1574--1580, 2021.

\bibitem{orvieto2023resurrecting}
Antonio Orvieto, Samuel~L Smith, Albert Gu, Anushan Fernando, Caglar Gulcehre, Razvan Pascanu, and Soham De.
\newblock Resurrecting recurrent neural networks for long sequences.
\newblock In {\em International Conference on Machine Learning}, pages 26670--26698. PMLR, 2023.

\bibitem{panigrahi2021mechanisms}
Anil Panigrahi and Bert~W O’Malley.
\newblock Mechanisms of enhancer action: the known and the unknown.
\newblock {\em Genome biology}, 22(1):108, 2021.

\bibitem{wang2022single}
Wang Qi, Yang Zhang, Bolei Zhang, Yao Fu, Xiaozhi Zhao, Jing Zhang, Ke Zuo, Yuexian Xing, Song Jiang, Zhaohui Qin, Erguang Li, Hongqian Guo, Hehe Liu, and Jingping Yang.
\newblock Single-cell chromatin accessibility landscape in kidney identifies additional cell-of-origin in heterogenous papillary renal cell carcinoma.
\newblock {\em Nature Communications}, 13(1):31, 2022.

\bibitem{quinlan2010bedtools}
Aaron~R Quinlan and Ira~M Hall.
\newblock Bedtools: a flexible suite of utilities for comparing genomic features.
\newblock {\em Bioinformatics}, 26(6):841--842, 2010.

\bibitem{ramasamy2023mediator}
Shyam Ramasamy, Abrar Aljahani, Magdalena~A Karpinska, TB~Ngoc Cao, Taras Velychko, J~Neos Cruz, Michael Lidschreiber, and A~Marieke Oudelaar.
\newblock The mediator complex regulates enhancer-promoter interactions.
\newblock {\em Nature Structural \& Molecular Biology}, 30(7):991--1000, 2023.

\bibitem{rood2024human}
Jennifer~E Rood, Samantha Wynne, Lucia Robson, Anna Hupalowska, John Randell, Sarah~A Teichmann, and Aviv Regev.
\newblock The human cell atlas from a cell census to a unified foundation model.
\newblock {\em Nature}, pages 1--2, 2024.

\bibitem{rubin2019coupled}
Adam~J Rubin, Kevin~R Parker, Ansuman~T Satpathy, Yanyan Qi, Beijing Wu, Alvin~J Ong, Maxwell~R Mumbach, Andrew~L Ji, Daniel~S Kim, Seung~Woo Cho, et~al.
\newblock Coupled single-cell crispr screening and epigenomic profiling reveals causal gene regulatory networks.
\newblock {\em Cell}, 176(1):361--376, 2019.

\bibitem{satpathy110k}
Ansuman~T Satpathy, Jeffrey~M Granja, Kathryn~E Yost, Yanyan Qi, Francesca Meschi, Geoffrey~P McDermott, Brett~N Olsen, Maxwell~R Mumbach, Sarah~E Pierce, M~Ryan Corces, et~al.
\newblock Massively parallel single-cell chromatin landscapes of human immune cell development and intratumoral t cell exhaustion.
\newblock {\em Nature biotechnology}, 37(8):925--936, 2019.

\bibitem{shtutman2011tumor}
Michael Shtutman, Mirza Baig, Elina Levina, Gregory Hurteau, Chang-uk Lim, Eugenia Broude, Mikhail Nikiforov, Timothy~T Harkins, C~Steven Carmack, Ye Ding, et~al.
\newblock Tumor-specific silencing of copz2 gene encoding coatomer protein complex subunit $\zeta$2 renders tumor cells dependent on its paralogous gene copz1.
\newblock {\em Proceedings of the National Academy of Sciences}, 108(30):12449--12454, 2011.

\bibitem{signac}
Tim Stuart, Avi Srivastava, Shaista Madad, Caleb Lareau, and Rahul Satija.
\newblock Single-cell chromatin state analysis with signac.
\newblock {\em Nature Methods}, 2021.
\newblock https://doi.org/10.1038/s41592-021-01282-5.

\bibitem{su2025scmultimap}
Chang Su, Dongsoo Lee, Peng Jin, and Jingfei Zhang.
\newblock scmultimap: Cell-type-specific mapping of enhancers and target genes from single-cell multimodal data.
\newblock {\em Nature Communications}, 16(1):3941, 2025.

\bibitem{sundaram2024single}
Laksshman Sundaram, Arvind Kumar, Matthew Zatzman, Adriana Salcedo, Neal Ravindra, Shadi Shams, Brian Louie, S Bagdatli, Shahab Sarmashghi, Hyo Choi, Wonyoung Choi, Kathryn Yost, Yanding Zhao, Jeffrey Granja, Toshinori Hinoue, D Hayes, Andrew Cherniack, Ina Felau, and Hui Shen.
\newblock Single-cell chromatin accessibility reveals malignant regulatory programs in primary human cancers.
\newblock {\em Science}, 385(6713):eadk9217, 2024.

\bibitem{tayyebi2024scalable}
Zakieh Tayyebi, Allison~R Pine, and Christina~S Leslie.
\newblock Scalable and unbiased sequence-informed embedding of single-cell atac-seq data with cellspace.
\newblock {\em Nature Methods}, 21(6):1014--1022, 2024.

\bibitem{terekhanova2023epigenetic}
Nadezhda~V. Terekhanova, Alla Karpova, Wen-Wei Liang, Alexander Strzalkowski, Siqi Chen, Li Yize, Austin~N. Southard-Smith, Iglesia~Michael D., Michael~C. Wendl, Reyka~G. Jayasinghe, Liu Jingxian, Yizhe Song, Song Cao, Andrew Houston, Xiuting Liu, Matthew~A. Wyczalkowski, Rita Jui-Hsien Lu, Wagma Caravan, Andrew Shinkle, Nataly Naser~Al Deen, John~M. Herndon, Jacqueline Mudd, Cong Ma, Hirak Sarkar, Kazuhito Sato, Omar~M. Ibrahim, Chia-Kuei Mo, Sara~E. Chasnoff, Eduard Porta-Pardo, Jason~M. Held, Russell Pachynski, Julie~K. Schwarz, William~E. Gillanders, Albert~H. Kim, Ravi Vij, John~F. DiPersio, Sidharth~V. Puram, Milan~G. Chheda, Katherine~C. Fuh, David~G. DeNardo, Ryan~C. Fields, Feng Chen, Benjamin~J. Raphael, and Li Ding.
\newblock Epigenetic regulation during cancer transitions across 11 tumour types.
\newblock {\em Nature}, 623(7986):432--441, 2023.

\bibitem{geneformer}
Christina~V Theodoris, Ling Xiao, Anant Chopra, Mark~D Chaffin, Zeina~R Al~Sayed, Matthew~C Hill, Helene Mantineo, Elizabeth~M Brydon, Zexian Zeng, X~Shirley Liu, et~al.
\newblock Transfer learning enables predictions in network biology.
\newblock {\em Nature}, 618(7965):616--624, 2023.

\bibitem{to2024multi}
Ken To, Lijiang Fei, J~Patrick Pett, Kenny Roberts, Raphael Blain, Krzysztof Pola{\'n}ski, Tong Li, Nadav Yayon, Peng He, Chuan Xu, et~al.
\newblock A multi-omic atlas of human embryonic skeletal development.
\newblock {\em Nature}, 635(8039):657--667, 2024.

\bibitem{van2017neural}
Aaron Van Den~Oord, Oriol Vinyals, et~al.
\newblock Neural discrete representation learning.
\newblock {\em Advances in neural information processing systems}, 30, 2017.

\bibitem{vaswani2017attention}
A Vaswani.
\newblock Attention is all you need.
\newblock {\em Advances in Neural Information Processing Systems}, 2017.

\bibitem{wang23k}
Lin Wang, Husam Babikir, S{\"o}ren M{\"u}ller, Garima Yagnik, Karin Shamardani, Francisca Catalan, Gary Kohanbash, Beatriz Alvarado, Elizabeth Di~Lullo, Arnold Kriegstein, et~al.
\newblock The phenotypes of proliferating glioblastoma cells reside on a single axis of variation.
\newblock {\em Cancer discovery}, 9(12):1708--1719, 2019.

\bibitem{wen2022graph}
Hongzhi Wen, Jiayuan Ding, Wei Jin, Yiqi Wang, Yuying Xie, and Jiliang Tang.
\newblock Graph neural networks for multimodal single-cell data integration.
\newblock In {\em Proceedings of the 28th ACM SIGKDD conference on knowledge discovery and data mining}, pages 4153--4163, 2022.

\bibitem{wen2023cellplm}
Hongzhi Wen, Wenzhuo Tang, Xinnan Dai, Jiayuan Ding, Wei Jin, Yuying Xie, and Jiliang Tang.
\newblock Cellplm: pre-training of cell language model beyond single cells.
\newblock {\em bioRxiv}, pages 2023--10, 2023.

\bibitem{wu2021babel}
Kevin~E Wu, Kathryn~E Yost, Howard~Y Chang, and James Zou.
\newblock Babel enables cross-modality translation between multiomic profiles at single-cell resolution.
\newblock {\em Proceedings of the National Academy of Sciences}, 118(15):e2023070118, 2021.

\bibitem{xiong2022online}
Lei Xiong, Kang Tian, Yuzhe Li, Weixi Ning, Xin Gao, and Qiangfeng~Cliff Zhang.
\newblock Online single-cell data integration through projecting heterogeneous datasets into a common cell-embedding space.
\newblock {\em Nature Communications}, 13(1):6118, 2022.

\bibitem{xiong2019scale}
Lei Xiong, Kui Xu, Kang Tian, Yanqiu Shao, Lei Tang, Ge Gao, Michael Zhang, Tao Jiang, and Qiangfeng~Cliff Zhang.
\newblock Scale method for single-cell atac-seq analysis via latent feature extraction.
\newblock {\em Nature communications}, 10(1):4576, 2019.

\bibitem{scanvi}
Chenling Xu, Romain Lopez, Edouard Mehlman, Jeffrey Regier, Michael~I Jordan, and Nir Yosef.
\newblock Probabilistic harmonization and annotation of single-cell transcriptomics data with deep generative models.
\newblock {\em Molecular systems biology}, 17(1):e9620, 2021.

\bibitem{xu12k}
Kun Xu, Wenwen Zhang, Cong Wang, Longfei Hu, Runtian Wang, Cenzhu Wang, Lin Tang, Guohua Zhou, Bingjie Zou, Hui Xie, et~al.
\newblock Integrative analyses of scrna-seq and scatac-seq reveal cxcl14 as a key regulator of lymph node metastasis in breast cancer.
\newblock {\em Human Molecular Genetics}, 30(5):370--380, 2021.

\bibitem{xu2024sstmultiscalehybridmambatransformer}
Xiongxiao Xu, Canyu Chen, Yueqing Liang, Baixiang Huang, Guangji Bai, Liang Zhao, and Kai Shu.
\newblock Sst: Multi-scale hybrid mamba-transformer experts for long-short range time series forecasting, 2024.

\bibitem{yang2022scbert}
Fan Yang, Wenchuan Wang, Fang Wang, Yuan Fang, Duyu Tang, Junzhou Huang, Hui Lu, and Jianhua Yao.
\newblock scbert as a large-scale pretrained deep language model for cell type annotation of single-cell rna-seq data.
\newblock {\em Nature Machine Intelligence}, 4(10):852--866, 2022.

\bibitem{yang2021multi}
Karren~Dai Yang, Anastasiya Belyaeva, Saradha Venkatachalapathy, Karthik Damodaran, Abigail Katcoff, Adityanarayanan Radhakrishnan, GV Shivashankar, and Caroline Uhler.
\newblock Multi-domain translation between single-cell imaging and sequencing data using autoencoders.
\newblock {\em Nature communications}, 12(1):31, 2021.

\bibitem{yang2024pan}
Yu Yang, Xueyan Chen, Jieying Pan, Huiheng Ning, Yaojun Zhang, Yufei Bo, Xianwen Ren, Jiesheng Li, Shishang Qin, Dongfang Wang, Min-Min Chen, and Zemin Zhang.
\newblock Pan-cancer single-cell dissection reveals phenotypically distinct b cell subtypes.
\newblock {\em Cell}, 187(17):4790--4811.e22, 2024.

\bibitem{yu2024robustifying}
Annan Yu, Arnur Nigmetov, Dmitriy Morozov, Michael~W. Mahoney, and N.~Benjamin Erichson.
\newblock Robustifying state-space models for long sequences via approximate diagonalization.
\newblock In {\em International Conference on Learning Representations}, 2024.

\bibitem{yuan2022scbasset}
Han Yuan and David~R Kelley.
\newblock scbasset: sequence-based modeling of single-cell atac-seq using convolutional neural networks.
\newblock {\em Nature Methods}, 19(9):1088--1096, 2022.

\bibitem{zeng2024deciphering}
Yuansong Zeng, Mai Luo, Ningyuan Shangguan, Peiyu Shi, Junxi Feng, Jin Xu, Ken Chen, Yutong Lu, Weijiang Yu, and Yuedong Yang.
\newblock Deciphering cell types by integrating scatac-seq data with genome sequences.
\newblock {\em Nature Computational Science}, 4(4):285--298, 2024.

\bibitem{zhang2021single}
Kai Zhang, James~D Hocker, Michael Miller, Xiaomeng Hou, Joshua Chiou, Olivier~B Poirion, Yunjiang Qiu, Yang~E Li, Kyle~J Gaulton, Allen Wang, et~al.
\newblock A single-cell atlas of chromatin accessibility in the human genome.
\newblock {\em Cell}, 184(24):5985--6001, 2021.

\bibitem{snapatac}
Kai Zhang, Nathan~R Zemke, Ethan~J Armand, and Bing Ren.
\newblock A fast, scalable and versatile tool for analysis of single-cell omics data.
\newblock {\em Nature methods}, 21(2):217--227, 2024.

\bibitem{script}
Yu Zhang, Baole Wen, Yifeng Jiao, Yuchen Liu, Xin Guo, Yushuai Wu, Jiyang Li, Limei Han, Yinghui Xu, Xin Gao, et~al.
\newblock Script: Predicting single-cell long-range cis-regulation based on pretrained graph attention networks.
\newblock {\em Advanced Science}, page e05021, 2025.

\bibitem{zhu2023multi}
Kaiyi Zhu, Jaroslav Bendl, Samir Rahman, James~M Vicari, Claire Coleman, Tereza Clarence, Ovaun Latouche, Nadejda~M Tsankova, Aiqun Li, Kristen~J Brennand, et~al.
\newblock Multi-omic profiling of the developing human cerebral cortex at the single-cell level.
\newblock {\em Science advances}, 9(41):eadg3754, 2023.

\bibitem{zhu2023single}
Qionghua Zhu, Xin Zhao, Yuanhang Zhang, Yanping Li, Shang Liu, Jingxuan Han, Zhiyuan Sun, Chunqing Wang, Daqi Deng, Shanshan Wang, et~al.
\newblock Single cell multi-omics reveal intra-cell-line heterogeneity across human cancer cell lines.
\newblock {\em Nature Communications}, 14(1):8170, 2023.

\bibitem{ziffra77k}
Ryan~S Ziffra, Chang~N Kim, Jayden~M Ross, Amy Wilfert, Tychele~N Turner, Maximilian Haeussler, Alex~M Casella, Pawel~F Przytycki, Kathleen~C Keough, David Shin, et~al.
\newblock Single-cell epigenomics reveals mechanisms of human cortical development.
\newblock {\em Nature}, 598(7879):205--213, 2021.

\end{thebibliography}
}

\appendix

\newpage

\section{Discussion of comparing OCR tokenization with other peak-calling algorithms}

Traditional scATAC-seq pipelines often rely on a fixed OCR vocabulary (e.g., from a reference atlas) to harmonize datasets, which ensures alignment consistency but introduces two major limitations: (1) a loss of OCR length information, which encodes accessibility strength and resolution, and (2) reduced generalizability, as tissue-specific or novel OCRs are often excluded. ChromFound addresses these issues through a genomic-coordinate–based tokenization strategy that encodes each OCR by its actual genomic start and end positions using sinusoidal positional embeddings (Section 3.1). This position-aware representation naturally accommodates variable-length OCRs and preserves their spatial relationships, enabling the model to capture regulatory heterogeneity across tissues.

We perform an experiment to directly compare ChromFound with VAE-based models trained on reference-peak-aligned data. Specifically, we adopt the cPeaks reference set~\cite{cpeaks}, which defines 1,657,194 peaks across the human genome. We use a 20,000-cell PBMC subset (same as in Table~\ref{ablation_study_table}, Row 6) as the training data. After mapping peaks to the cPeaks reference and filtering out peaks and cells with all-zero values, we obtain a training set of 306,784 peaks and approximately 18,000 cells.

We train four VAE-based baselines~\cite{xiong2019scale,xiong2022online,peakvi,poissonvi} on the aligned data and compare their performance on zero-shot cell clustering (Table~\ref{clustering_table}) against ChromFound trained on the same aligned data. Evaluation metrics and datasets follow Table\ref{clustering_table}, all aligned to 306,784 peaks. Results are summarized in Table~\ref{cpeaks}.

This comparison yields two key observations. First, reference-based alignment substantially limits cross-tissue generalization: only 30\% of peaks in retina, 35\% in cortex, and 70\% in heart overlap with the cPeaks-aligned PBMC training peaks, leading to markedly lower performance of the baseline VAEs on unseen tissues. Second, ChromFound, though robust even when trained on the reference-aligned data, performs best when using its native OCRs, suggesting that fixed-peak harmonization fails to capture novel or shifted OCRs arising from tissue heterogeneity, disease-specific variation, or batch effects.

\begin{table}
\caption{Results of VAE-based models with cPeaks-aligned training against ChromFound. Result style: \textbf{best}, \uline{second best}}
\centering
\renewcommand{\arraystretch}{1.0}
\begin{tabular}{@{}l|l|c|c|cccc@{}}
\toprule
\textbf{Dataset}                      & \textbf{Model} & \textbf{VAE} & \textbf{cPeaks} & \textbf{ARI}($\uparrow$)                                          & \textbf{FMI}($\uparrow$)                                          & \textbf{NMI}($\uparrow$)                                          & \textbf{AMI}($\uparrow$)                                          \\ \midrule
                                      & SCALE~\cite{xiong2019scale} & \ding{52}         & \ding{52}                                                                                        & 0.2932                                                & 0.5050                                                & 0.4127                                                & 0.4081                                                \\
                                      & SCALEX~\cite{xiong2022online} & \ding{52}         & \ding{52}                                                                                        & 0.4230                                                & 0.6078                                                & 0.5543                                                & 0.5513                                                \\
                                      & peakVI~\cite{peakvi} & \ding{52}         & \ding{52}                                                                                        & 0.3690                                                & 0.5643                                                & 0.5023                                                & 0.4983                                                \\
                                      & poissonVI~\cite{poissonvi} & \ding{52}         & \ding{52}                                                                                        & 0.3364                                                & 0.5408                                                & 0.4636                                                & 0.4594                                                \\
                                      & ChromFound  &         & \ding{52}                                                                                        & \uline{0.5858}                                                & \uline{0.7244}                                                & \uline{0.7519}                                                & \uline{0.7499}                                                \\
\multirow{-6}{*}{Cortex(batch 1)}     & ChromFound &         &                                                                                        & \textbf{0.6475} & \textbf{0.7503} & \textbf{0.7701} & \textbf{0.7689} \\ \midrule
                                      & SCALE~\cite{xiong2019scale}          & \ding{52}         & \ding{52}                                                                                        & 0.2484                                                & 0.4052                                                & 0.3657                                                & 0.3609                                                \\
                                      & SCALEX~\cite{xiong2022online}        & \ding{52}         & \ding{52}                                                                                        & 0.4481                                                & 0.5679                                                & 0.5742                                                & 0.5710                                                \\
                                      & peakVI~\cite{peakvi}         & \ding{52}         & \ding{52}                                                                                        & 0.3610                                                & 0.5035                                                & 0.4876                                                & 0.4835                                                \\
                                      & poissonVI~\cite{poissonvi}      & \ding{52}         & \ding{52}                                                                                        & 0.2820                                                & 0.4374                                                & 0.4104                                                & 0.4058                                                \\
                                      & ChromFound     &         & \ding{52}                                                                                        & \uline{0.5751}                                                & \uline{0.6720}                                                & \uline{0.7145}                                                & \uline{0.7123}                                                \\
\multirow{-6}{*}{Cortex(batch 2)}     & ChromFound     &         &                                                                                        & \textbf{0.6092} & \textbf{0.7015} & \textbf{0.7276} & \textbf{0.7259} \\ \midrule
                                      & SCALE~\cite{xiong2019scale}          & \ding{52}         & \ding{52}                                                                                        & 0.3543                                                & 0.4817                                                & 0.5228                                                & 0.5195                                                \\
                                      & SCALEX~\cite{xiong2022online}         & \ding{52}         & \ding{52}                                                                                        & 0.2795                                                & 0.4150                                                & 0.4787                                                & 0.4752                                                \\
                                      & peakVI~\cite{peakvi}         & \ding{52}         & \ding{52}                                                                                        & 0.3023                                                & 0.4437                                                & 0.4816                                                & 0.4779                                                \\
                                      & poissonVI~\cite{poissonvi}      & \ding{52}         & \ding{52}                                                                                        & 0.3778                                                & 0.5025                                                & 0.5446                                                & 0.5414                                                \\
                                      & ChromFound     &         & \ding{52}                                                                                        & \uline{0.5400}                                                & \uline{0.6347}                                                & \uline{0.6588}                                                & \uline{0.6567}                                                \\
\multirow{-6}{*}{Heart(av 3)}         & ChromFound     &         &                                                                                        & \textbf{0.5642} & \textbf{0.6601} & \textbf{0.6981} & \textbf{0.6953} \\ \midrule
                                      & SCALE~\cite{xiong2019scale}          & \ding{52}         & \ding{52}                                                                                        & 0.4051                                                & 0.6506                                                & 0.4461                                                & 0.4453                                                \\
                                      & SCALEX~\cite{xiong2022online}         & \ding{52}         & \ding{52}                                                                                        & 0.3013                                                & 0.5685                                                & 0.4199                                                & 0.4191                                                \\
                                      & peakVI~\cite{peakvi}         & \ding{52}         & \ding{52}                                                                                        & 0.3966                                                & 0.6080                                                & 0.5018                                                & 0.5011                                                \\
                                      & poissonVI~\cite{poissonvi}      & \ding{52}         & \ding{52}                                                                                        & 0.3939                                                & 0.5997                                                & 0.5010                                                & 0.5003                                                \\
                                      & ChromFound     &         & \ding{52}                                                                                        & \uline{0.6150}                                                & \uline{0.7695}                                                & \uline{0.6968}                                                & \uline{0.6963}                                                \\
\multirow{-6}{*}{Heart(av 10)}        & ChromFound     &         &                                                                                        & \textbf{0.6432} & \textbf{0.7869} & \textbf{0.7220} & \textbf{0.7223} \\ \midrule
                                      & SCALE~\cite{xiong2019scale}          & \ding{52}         & \ding{52}                                                                                        & 0.3065                                                & 0.5771                                                & 0.4447                                                & 0.4440                                                \\
                                      & SCALEX~\cite{xiong2022online}         & \ding{52}         & \ding{52}                                                                                        & 0.2430                                                & 0.5149                                                & 0.4127                                                & 0.4117                                                \\
                                      & peakVI~\cite{peakvi}         & \ding{52}         & \ding{52}                                                                                        & 0.2514                                                & 0.5260                                                & 0.4651                                                & 0.4644                                                \\
                                      & poissonVI~\cite{poissonvi}      & \ding{52}         & \ding{52}                                                                                        & 0.2347                                                & 0.5122                                                & 0.4063                                                & 0.4055                                                \\
                                      & ChromFound     &         & \ding{52}                                                                                        & \uline{0.4888}                                                & \uline{0.7081}                                                & \uline{0.7241}                                                & \uline{0.7238}                                                \\
\multirow{-6}{*}{Retina(D026\_13)}    & ChromFound     &         &                                                                                        & \textbf{0.5802} & \textbf{0.7549} & \textbf{0.7410} & \textbf{0.7403} \\ \midrule
                                      & SCALE~\cite{xiong2019scale}          & \ding{52}         & \ding{52}                                                                                        & 0.3143                                                & 0.5092                                                & 0.4220                                                & 0.4210                                                \\
                                      & SCALEX~\cite{xiong2022online}         & \ding{52}         & \ding{52}                                                                                        & 0.2490                                                & 0.4542                                                & 0.3920                                                & 0.3909                                                \\
                                      & peakVI~\cite{peakvi}         & \ding{52}         & \ding{52}                                                                                        & 0.2664                                                & 0.4693                                                & 0.4745                                                & 0.4736                                                \\
                                      & poissonVI~\cite{poissonvi}      & \ding{52}         & \ding{52}                                                                                        & 0.2396                                                & 0.4473                                                & 0.3937                                                & 0.3926                                                \\
                                      & ChromFound     &         & \ding{52}                         & \uline{0.5368}                                                & \uline{0.6839}                                                & \uline{0.7664}                                                & \uline{0.7660}                                                \\
\multirow{-6}{*}{Retina(D19D008)}     & ChromFound     &         &                                                                                        & \textbf{0.5956} & \textbf{0.7202} & \textbf{0.7832} & \textbf{0.7868} \\ \midrule
                                      & SCALE~\cite{xiong2019scale}          & \ding{52}         & \ding{52}                                                                                       & 0.5769                                                & 0.6756                                                & 0.7066                                                & 0.7057                                                \\
                                      & SCALEX~\cite{xiong2022online}         & \ding{52}         & \ding{52}                                                                                        & 0.5718                                                & 0.6514                                                & 0.7205                                                & 0.7196                                                \\
                                      & peakVI~\cite{peakvi}         & \ding{52}         & \ding{52}                                                                                        & 0.5306                                                & 0.6249                                                & 0.7057                                                & 0.7046                                                \\
                                      & poissonVI~\cite{poissonvi}      & \ding{52}         & \ding{52}                                                                                        & 0.5593                                                & 0.6479                                                & 0.6742                                                & 0.6732                                                \\
                                      & ChromFound     &         & \ding{52}                                                                                        & \uline{0.5999}                                                & \uline{0.6827}                                                & \uline{0.7434}                                                & \uline{0.7424}                                                \\
\multirow{-6}{*}{PBMC(VIB\_10xv1\_1)} & ChromFound     &         &                                                                                        & \textbf{0.6142} & \textbf{0.6995} & \textbf{0.7354} & \textbf{0.7345} \\ \midrule
                                      & SCALE~\cite{xiong2019scale}          & \ding{52}         & \ding{52}                                                                                        & 0.4031                                                & 0.5797                                                & 0.5380                                                & 0.5374                                                \\
                                      & SCALEX~\cite{xiong2022online}         & \ding{52}         & \ding{52}                                                                                        & 0.3326                                                & 0.5430                                                & 0.5465                                                & 0.5458                                                \\
                                      & peakVI~\cite{peakvi}         & \ding{52}         & \ding{52}                                                                                        & 0.3830                                                & 0.5804                                                & 0.5737                                                & 0.5731                                                \\
                                      & poissonVI~\cite{poissonvi}      & \ding{52}         & \ding{52}                                                                                        & 0.3959                                                & 0.5917                                                & 0.5509                                                & 0.5503                                                \\
                                      & ChromFound     &         & \ding{52}                                                                                        & \uline{0.4347}                                                & \uline{0.6237}                                                & \uline{0.5867}                                                & \uline{0.5861}                                                \\
\multirow{-6}{*}{PBMC(BIO\_ddseq\_1)} & ChromFound     &         &                                                                                       & \textbf{0.4579} & \textbf{0.6421} & \textbf{0.5902} & \textbf{0.5910} \\ \bottomrule
\end{tabular}
\label{cpeaks}
\end{table}

\section{Experiments details}
\label{preliminary_exp}

\subsection{Cell clustering}
\label{sec:cell_clustering_add}
Since ChromFound is configured in a zero-shot mode for the cell clustering task, no additional fine-tuning architecture or training parameters need to be specified. We directly propagate the data through the encoder structure of ChromFound, extracting the encoder output as the final representation. After performing a pooling layer on the hidden dimension, we apply Principal Component Analysis (PCA) to reduce the dimensionality of the output to 50. Based on this, we compute the metrics for cell clustering and generate the UMAP clustering visualization. The benchmark methods are implemented with the default parameters provided by their source codes.

We compute the inference speed of ChromFound on a node equipped with 32 CPU cores, 256 GB of memory, and one NVIDIA A100 GPU. The input scATAC-seq dataset with 232,354 OCRs. The maximum CPU memory usage is 23.9 GB and the maximum GPU memory consumption is 12.7 GB out of 80 GB available. The inference speed with batch size 4 is 0.97 seconds per batch.

\subsection{Cell type annotation}
\label{cell_type_annotation_details}
Building upon our four-layer pretrained model, we extract the encoder output as a general-purpose representation for downstream cell type annotation. To align the model with this task, we incorporate additional fine-tuning layers. Specifically, the hidden dimension produced by ChromFound’s encoder is expanded from 128 to 256 before being projected down to a single dimension, resulting in a pooling tensor. Subsequently, additional multilayer perceptron (MLP) layers are employed to predict the logits corresponding to cell types. We freeze the decoder parameters from pretraining while enabling gradient flow through the rest of the pretrained backbone so that it can adapt to the new classification objective. Dropout mitigates overfitting by randomly zeroing hidden units, while LayerNorm helps stabilize and accelerate fine-tuning. The benchmark methods are implemented with the default parameters provided by their source codes.

For experimental settings, we divide the training data into 90\% for training and 10\% for validation. We train for 20 epochs using the AdamW optimizer with an initial learning rate of $5\times 10^{-4}$ and a linear warm-up schedule of 50 steps. The best model is chosen based on validation set performance and evaluated on the test set to obtain the reported accuracy and macro F1 score. All experiments are conducted on a single machine equipped with four NVIDIA A100 GPUs.

\subsection{Cross-omics prediction}
\label{modality_prediction_appendix}
We employ a two-layer Multilayer Perceptron (MLP) to perform the modality prediction task, with the hidden layer size set to 1024. The learning rate is configured at 1e-4, and AdamW is utilized as the optimizer. A total of 10 epochs are conducted for training. All benchmark methods are implemented with the default parameters in the DANCE package~\cite{ding2022dance}. All experiments are conducted on a single machine equipped with four NVIDIA A100 GPUs. We divide the whole dataset into 80\% for training, 10\% for evaluation, and 10\% for testing.

\subsection{Biological application}
\label{biological_app_appendix}
To validate the biological relevance of predicted enhancer-gene interactions, we simulate enhancer knockdown using our cross-omics model trained on the Zhu11K dataset~\cite{zhu2023single}. Specifically, we select 141 K562 cells from the test split and iteratively set the accessibility of each candidate enhancer to zero, mimicking CRISPRi-mediated repression. The model then infers post-perturbation gene expression, and the change relative to the unperturbed state is computed and averaged across all cells. The absolute magnitude of this change indicates the likelihood of a regulatory interaction, while the signed change reflects its direction and strength.

We focus on two cancer-related genes, COPZ1 and HNRNPA1, which are extensively characterized in the Fulco4K CRISPRi dataset~\cite{zhu2023single}. Each gene is associated with 117 candidate enhancers, among which 6 (COPZ1) and 10 (HNRNPA1) are experimentally validated.  To resolve discrepancies in genomic coordinates between Fulco4K~\cite{fulco2019activity} and Zhu11K~\cite{zhu2023single}, we perform enhancer mapping using bedtools~\cite{quinlan2010bedtools}. In total, the simulation spans 230,819 enhancers and 17,476 genes.

To evaluate performance, we compute the area under the ROC curve seen in Table~\ref{roc_curves} between the absolute expression change and the binary labels of enhancer-gene links tested with a statistical significance on gene expression at a false discovery rate (FDR) < 0.05~\cite{fulco2019activity}. ChromFound achieves the highest accuracy, with AUCs of 0.77 for COPZ1 and 0.61 for HNRNPA1, demonstrating its ability to identify functional regulatory links. Additionally, we calculate the Pearson correlation between signed responses and CRISPRi-measured quantitative effects plotted in Table~\ref{perturbation}.

The Benchmark methods BABEL~\cite{wu2021babel} and CMAE~\cite{yang2021multi} are included for comparison due to their strong cross-omics performance on Zhu11K. However, both rely on filtering the input scATAC-seq profiles to the top 10,000 highly variable OCRs~\cite{ding2022dance}, which excludes the majority of candidate enhancers from Fulco4K. As a result, these methods often fail to produce meaningful predictions for enhancer perturbations, yielding near-zero expression changes in most cases. This limitation is evident in both the ROC and scatter plots, underscoring the advantage of ChromFound’s genome-wide modeling in capturing fine-grained regulatory effects at scale.

\section{Details of datasets}
\label{datasets_details}
We collect a large-scale human scATAC-seq dataset that includes more than 2.65 million cells and 1.75 trillion tokens. ChromFound for pretraining is based on two fundamental datasets, the human atlas~\cite{zhang2021single} and the fetal atlas~\cite{domcke2020human}, which contribute 1.32 million cells spanning multiple human organs and serve as the primary sources for pretraining ChromFound. We accept the CRCh38 as reference genome across all datasets to maintain uniformity, converting datasets originally in hg19 to hg38 when necessary. The resources of all datasets are detailed in Table~\ref{tab:dataset_info}.

\section{Pre-processing of scATAC-seq data}
\label{sec:data_processing}
We develop a preprocessing pipeline for scATAC-seq data to address sparsity and high dimensionality, ensuring high-quality cells and OCRs for model pretraining and downstream analyses. Preprocessing is applied after train/test splitting to avoid information leakage, and all datasets are stored in \texttt{.h5ad} format with genomic OCR coordinates in the \texttt{var} schema.

\textbf{Quality Control}: We filter out cells with non-zero counts below 1,000 or above 60,000 and remove OCRs with non-zero counts present in less than 5\% of all cell types to preserve heterogeneity among rare populations. For unlabeled datasets, pseudo labels are inferred using TF-IDF followed by LSI for OCR filtering.

\textbf{Cell aggregation}: To mitigate sparsity, we aggregate neighboring cells identified through Latent Semantic Indexing (LSI) and cosine similarity. LSI extracts $n$ components after removing technical noise (the first component), and cosine similarity determines the 10 nearest neighbors for aggregation.

\textbf{Normalization and Log Transformation}: To stabilize variance and scale the data, we apply total count normalization followed by log transformation.

\begin{table}[H]
\caption{Dataset Information}
\label{tab:dataset_info}
\centering
\renewcommand{\arraystretch}{0.97}
\footnotesize 
\begin{adjustbox}{max width=\textwidth} 
\begin{tabular}{>{\centering\arraybackslash}m{2.8cm} >{\centering\arraybackslash}m{2cm} >{\centering\arraybackslash}m{1.8cm} >{\centering\arraybackslash}m{1.4cm} >{\centering\arraybackslash}m{1.4cm} >{\centering\arraybackslash}m{0.9cm} >{\centering\arraybackslash}m{2cm}}
\toprule
\rowcolor{headerblue}
\multicolumn{1}{c}{\textbf{Dataset}} & \multicolumn{1}{c}{\textbf{Tissue}} & \multicolumn{1}{c}{\textbf{Disease}} & \multicolumn{1}{c}{\textbf{Cell Numbers}} & \multicolumn{1}{c}{\textbf{OCR Numbers}} & \multicolumn{1}{c}{\textbf{Genome}} & \multicolumn{1}{c}{\textbf{GSE/GSM ID}} \\
\midrule
\makecell{HumanAtlas~\cite{zhang2021single}} & Atlas & Health & 615,998 & 1,154,611 & hg38 & GSE184462 \\ \midrule
\makecell{FetalAtlas~\cite{domcke2020human}} & Atlas & Health & 707,043 & 1,154,646 & hg38 & GSE149683 \\ \midrule
\makecell{PBMCBenchmark~\cite{de2024systematic}} & Peripheral Blood & Health & 169,047 & 412,490 & hg38 & GSE194028 \\ \midrule
\makecell{Cortex130K~\cite{cortex63k}} & Cortex & Alzheimer’s disease & 130,419 & 219,070 & hg38 & GSE174367 \\ \midrule
\makecell{Cortex2K~\cite{cortex2k}} & Cortex & Health & 2,174 & 292,156 & hg38 & GSE174226 \\ \midrule
\makecell{Zhu45K~\cite{zhu2023multi}} & Cortex & Health & 45,549 & 304,034 & hg38 & GSE204684 \\ \midrule
\makecell{PBMC9K~\cite{pbmc10k}} & Peripheral Blood & Leukemia & 9,215 & 108,344 & hg38 & GSE139369 \\ \midrule
\makecell{BMMC11K~\cite{pbmc10k}} & Bone Marrow & Health & 11,384 & 452,004 & hg19 & GSE194122 \\ \midrule
\makecell{Rubin~\cite{rubin2019coupled}} & Epidermis & Health & 288 & 94,633 & hg19 & GSE116428 \\ \midrule
\multirow{2}{*}{\centering\makecell{Xu12K~\cite{xu12k}}} & \multirow{2}{*}{\centering Breast} & \multirow{2}{*}{\centering Breast cancer} & 7,202 & 87,851 & \multirow{2}{*}{\centering hg38} & GSM4798906 \\ 
& & & 5,642 & 76,850 & & GSM4798907 \\ \midrule
\multirow{6}{*}{\centering\makecell{Wang23K~\cite{wang23k}}} & \multirow{6}{*}{\centering Brain} & \multirow{6}{*}{\centering Glioma} & 6,284 & 79,730 & \multirow{6}{*}{\centering hg38} & GSM4119513 \\
& & & 129 & 1,169 & & GSM4119514 \\
& & & 5,213 & 210,434 & & GSM4119515 \\
& & & 5,519 & 183,847 & & GSM4119516 \\
& & & 2,229 & 40,907 & & GSM4119517 \\
& & & 3,628 & 93,121 & & GSM4119518 \\ \midrule
\makecell{Buenrostro3K~\cite{buenrostro3k}} & Bone Marrow & Health & 2,953 & 491,437 & hg19 & GSE96769 \\ \midrule
\makecell{Corces0.5K~\cite{corces0.5k}} & \makecell{Bone Marrow/\\Peripheral Blood} & Leukemia & 576 & 590,650 & hg19 & GSE74310 \\ \midrule
\makecell{Corces70K~\cite{corces2020single}} & Brain & \makecell{Alzheimer's and\\Parkinson's diseases} & 70,631 & 444,747 & hg38 & GSE147672 \\ \midrule
\multirow{4}{*}{\centering\makecell{Satpathy110K~\cite{satpathy110k}}} & Bone Marrow & Health & 63,882 & 571,400 & \multirow{4}{*}{\centering hg19} & \multirow{4}{*}{\centering GSE129785} \\
& Peripheral Blood & Health & 4,146 & 238,616 & & \\
& Peripheral Blood & Health & 4,786 & 152,367 & & \\
& TME & Cancer & 37,818 & 580,789 & & \\ \midrule
\makecell{Ziffra77K~\cite{ziffra77k}} & Forebrain & Health & 77,354 & 459,953 & hg38 & GSE163018 \\ \midrule
\makecell{Kuppe139K~\cite{kuppe2022spatial}} & Heart & Myocardial infarction & 139,835 & 429,828 & hg38 & -- \\ \midrule
\makecell{Liang154K~\cite{liang2023multi}} & Retina & Health & 154,775 & 264,833 & hg19 & GSE226108 \\ \midrule
\makecell{To326K~\cite{to2024multi}} & Bone & Health & 326,532 & 530,751 & hg38 & -- \\ \midrule
\makecell{Zhu11K~\cite{zhu2023single}} & Cell Line & N/A & 11,632 & 258,044 & hg19 & GSE118912 \\ \midrule
\multirow{5}{*}{\centering\makecell{10x PBMC Datasets}} & \multirow{5}{*}{\centering Peripheral Blood} & \multirow{5}{*}{\centering Health} & 10,247 & 90,686 & \multirow{5}{*}{\centering hg19} & \href{https://www.10xgenomics.com/datasets/10-k-peripheral-blood-mononuclear-cells-pbm-cs-from-a-healthy-donor-next-gem-v-1-1-1-1-standard-1-1-0}{10k v1.1}\tnote{1} \\
& & & 9,688 & 144,023 & & \href{https://www.10xgenomics.com/datasets/10-k-peripheral-blood-mononuclear-cells-pbm-cs-from-a-healthy-donor-next-gem-v-1-1-1-1-standard-2-0-0}{10k v2.0}\tnote{2} \\
& & & 4,623 & 135,377 & & \href{https://www.10xgenomics.com/datasets/5-k-peripheral-blood-mononuclear-cells-pbm-cs-from-a-healthy-donor-next-gem-v-1-1-1-1-standard-2-0-0}{5k v2.0}\tnote{3} \\
& & & 1,004 & 82,579 & & \href{https://www.10xgenomics.com/datasets/1-k-peripheral-blood-mononuclear-cells-pbm-cs-from-a-healthy-donor-next-gem-v-1-1-1-1-standard-2-0-0}{1k v2.0}\tnote{4} \\
& & & 484 & 65,908 & & \href{https://www.10xgenomics.com/datasets/500-peripheral-blood-mononuclear-cells-pbm-cs-from-a-healthy-donor-next-gem-v-1-1-1-1-standard-2-0-0}{500 v2.0}\tnote{5} \\
\bottomrule
\end{tabular}
\end{adjustbox}
\end{table}

\section{Supplementary results of ablation study}
\subsection{Hyperparameter sensitivity}
We evaluate the hyperparameter sensitivity on the cell clustering task using PBMC169K (batch VIB\_10xv1\_1) dataset.

\textbf{Positional embedding temperature}: This parameter controls the frequency of sinusoidal position encodings and can be viewed as a proxy for "genomic resolution". As shown in Table~\ref{temp_table}, performance slightly drops when temp is within 10e3 to 10e5, and degrades more substantially when temp is too large, likely due to loss of relative position sensitivity.

\begin{table}[htbp]
\caption{Results of hyperparameter $temp$ sensitivity experiment.}
\centering
\label{temp_table}
\begin{tabular}{@{}c|cccc@{}}
\toprule
\textbf{$temp$} & \textbf{ARI($\uparrow$)}    & \textbf{NMI($\uparrow$)}    & \textbf{AMI($\uparrow$)}    & \textbf{FMI($\uparrow$)}    \\ \midrule
1000            & 0.6535          & 0.7764          & 0.7755          & 0.7259          \\
10000           & 0.6512          & 0.7709          & 0.7701          & 0.7243          \\
\textbf{100000 (Ours)}   & \textbf{0.6953} & \textbf{0.7860} & \textbf{0.7852} & \textbf{0.7601} \\
1000000         & 0.6036          & 0.7344          & 0.7334          & 0.6860          \\
10000000        & 0.5852          & 0.7269          & 0.7258          & 0.6714          \\ \bottomrule
\end{tabular}
\end{table}

\textbf{Mamba projection dimension}: This parameter controls the compression within the Mamba block. As shown in Table~\ref{mamba_proj_dim}, $D_{low}$ = 32 achieves a strong balance between performance and efficiency. Our choice of 32 is primarily motivated by the trade-off between performance and computational efficiency. Larger values lead to marginal gains but significantly increase FLOPs and memory, with $D_{low}$ = 128 exceeding the memory limits on A100 80G GPUs.

\begin{table}[htbp]
\caption{Results of Mamba projection dimension $D_{low}$ sensitivity experiment.}
\centering
\label{mamba_proj_dim}
\begin{tabular}{@{}c|c|c|c|c|c@{}}
\toprule
\textbf{$D_{low}$} & \textbf{\#Parameter} & \textbf{FLOPs (10e11)}               & \textbf{Inference Speed (s)} & \textbf{VRAM (GB)} & \textbf{ARI($\uparrow$)}    \\ \midrule
16                 & 422,593                  & 7.41          & 3.4027                              & 48.8                     & 0.6158          \\
\textbf{32 (Ours)} & \textbf{450,305}         & \textbf{7.89} & \textbf{3.4624}                     & \textbf{60.7}            & \textbf{0.6953} \\
64                 & 518,785                  & 9.09         & 4.0053                              & 72.3                     & 0.6927          \\
128                & 707,969                  & 12.4          & OOM                                 & OOM                      & OOM             \\ \bottomrule
\end{tabular}
\end{table}

\subsection{Comparison on Transformer-only architectures}
Training a vanilla Transformer on scATAC-seq inputs containing nearly one million peaks is practically infeasible due to its quadratic scaling in memory and computation. To approximate a pure Transformer baseline under practical settings, we train two representative single-cell foundation models, Geneformer and scGPT, on the same pretraining corpus with increasing peak lengths (4k–32k) and the same hyperparameters as WPSA. The results of comparison on the cell clustering task using the PBMC169K (batch VIB\_10xv1\_1) dataset are detailed in Table~\ref{transformer_only}.

\begin{table}[htbp]
\caption{Results of comparison on Transformer-only architectures.}
\centering
\begin{tabular}{@{}l|c|c@{}}
\toprule
\textbf{Method}     & \textbf{OCRs Length} & \textbf{ARI($\uparrow$)}    \\ \midrule
Geneformer          & 4096                 & 0.0451          \\
                    & 16384                & 0.0457          \\
                    & 32768                & 0.0460          \\ \midrule
scGPT               & 4096                 & 0.3075          \\
                    & 16384                & 0.3774          \\
                    & 32768                & 0.3868          \\ \midrule
\textbf{ChromFound} & \textbf{440,000}     & \textbf{0.6953} \\ \bottomrule
\end{tabular}
\label{transformer_only}
\end{table}

\section{Supplementary results of denoising batch effect}
\begin{figure}
  \includegraphics[width=\textwidth]{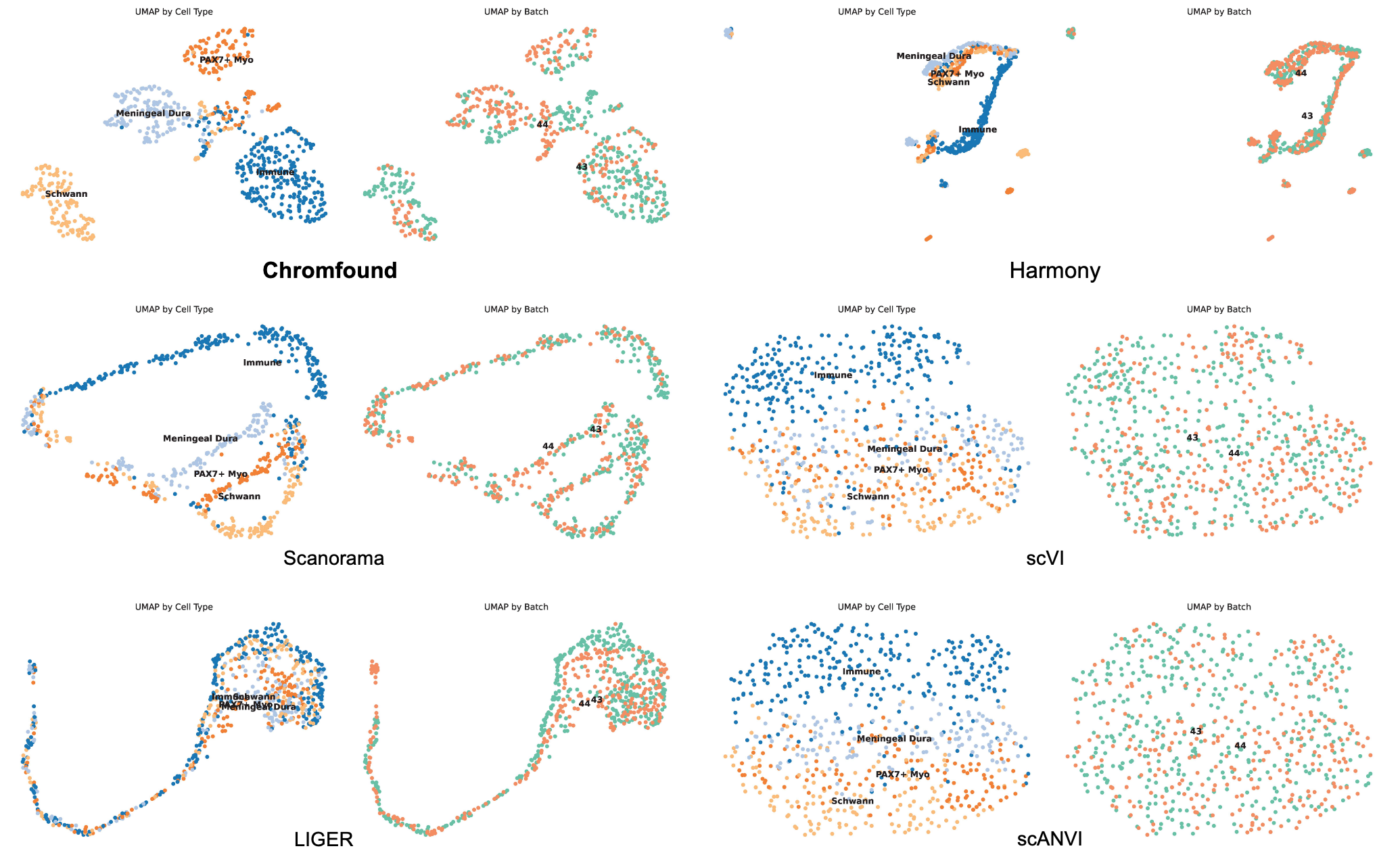}
  \caption{
  The plots illustrate ChromFound's superior performance in denoising batch effect on \textbf{Bone} To326K~\cite{to2024multi}.}
  \label{bone_batch_removal_compasion}
\end{figure}

\begin{figure}
  \includegraphics[width=\textwidth]{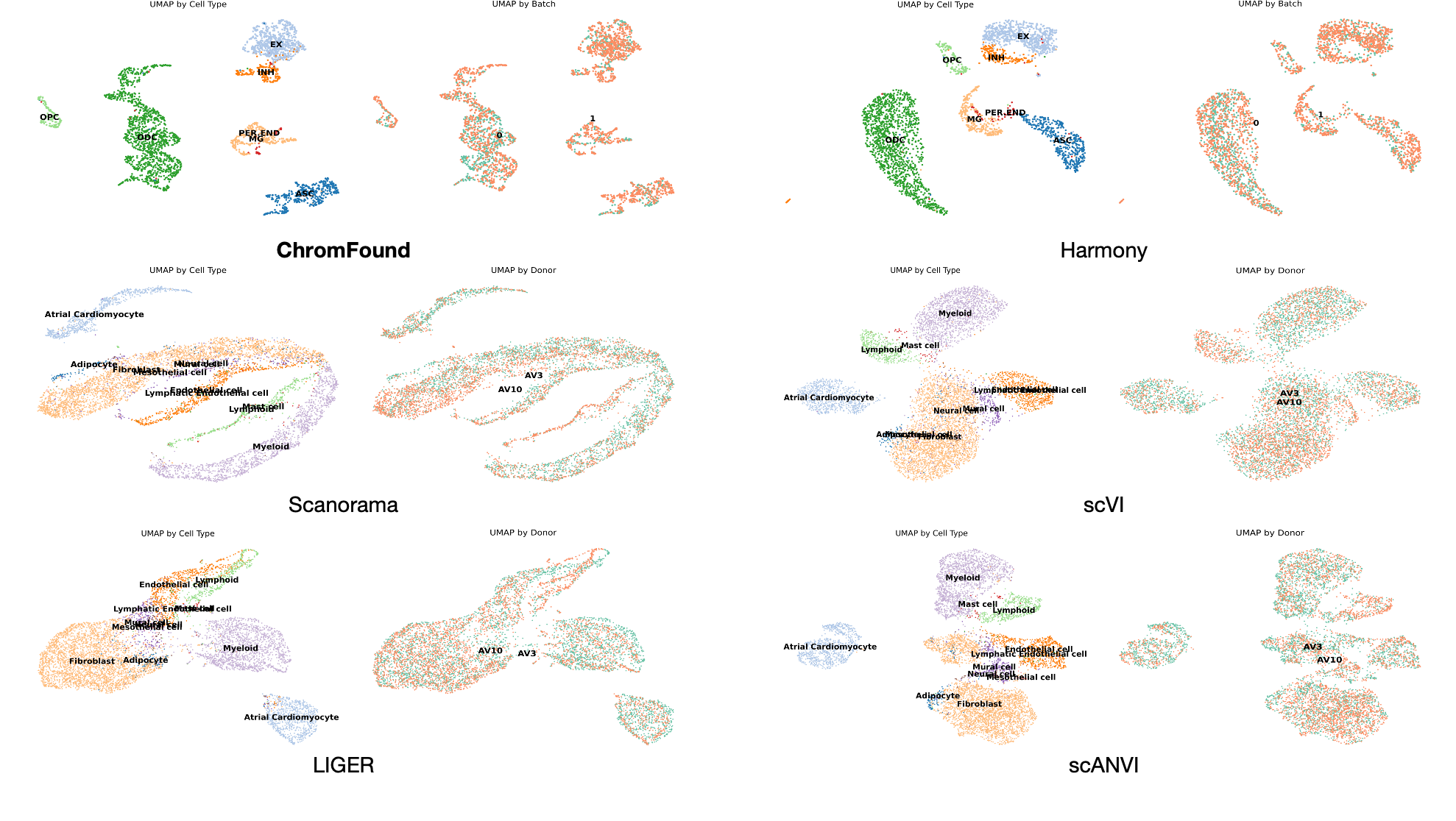}
  \caption{
  The plots illustrate ChromFound's superior performance in denoising batch effect on \textbf{Heart} Kuppel139K~\cite{kuppe2022spatial}.}
  \label{heart_batch_removal_compasion}
\end{figure}

\begin{figure}
  \includegraphics[width=\textwidth]{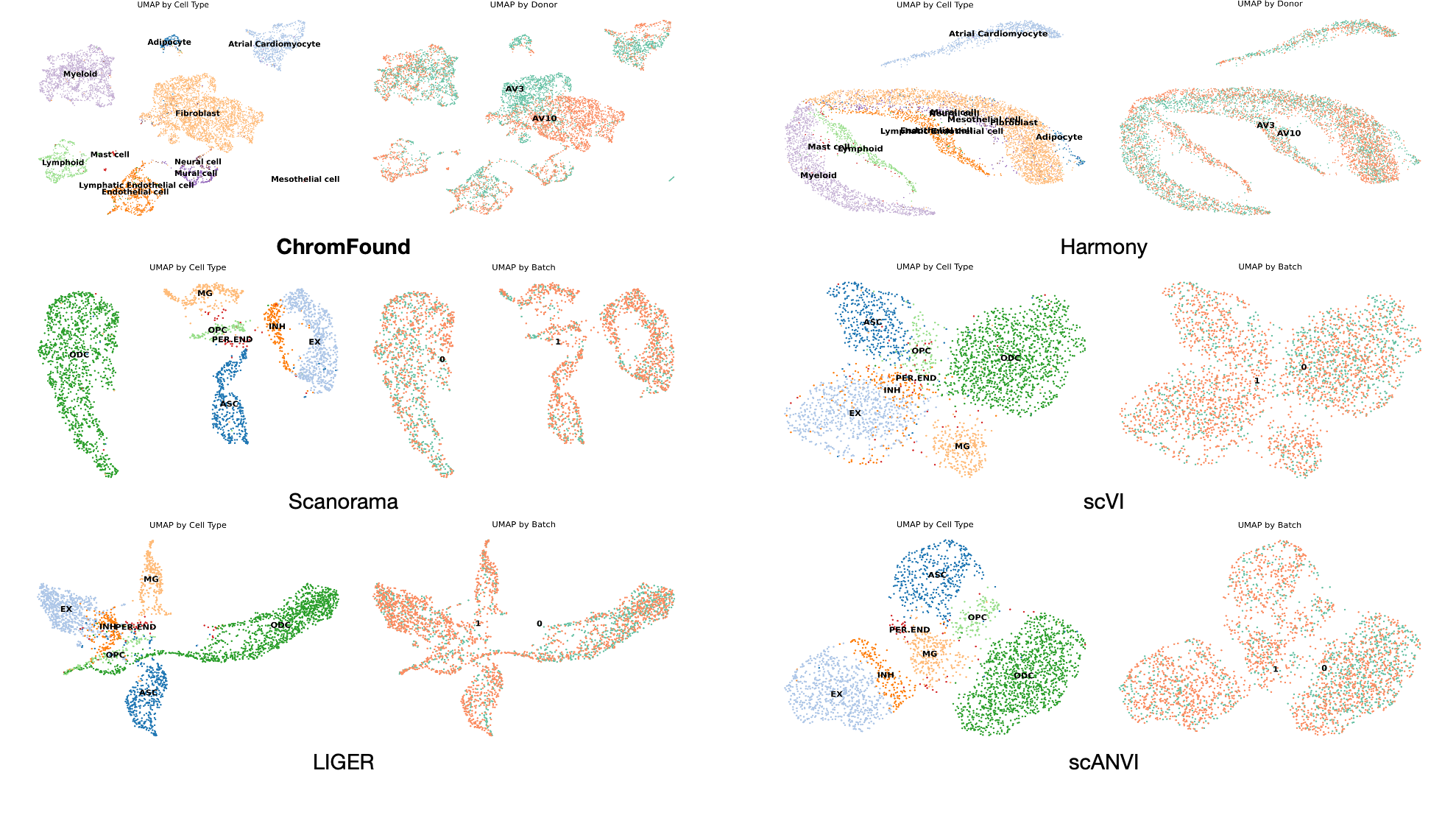}
  \caption{
  The plots illustrate ChromFound's superior performance in denoising batch effect on \textbf{Cortex} Morabitol130K~\cite{cortex63k}.}
  \label{cortex_batch_removal_compasion}
\end{figure}

\begin{figure}
  \includegraphics[width=\textwidth]{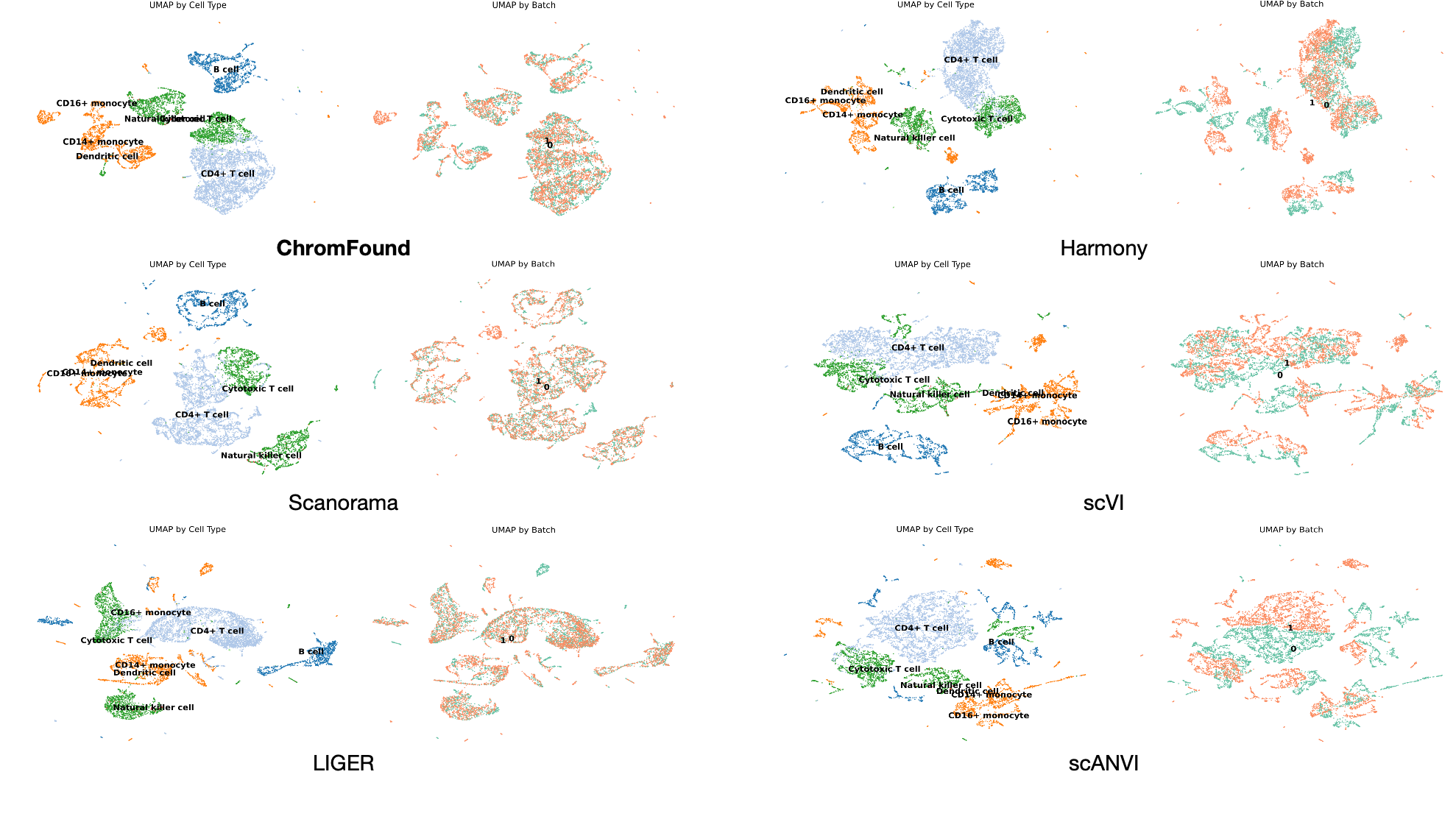}
  \caption{
  The plots illustrate ChromFound's superior performance in denoising batch effect on \textbf{PBMC} 169K~\cite{de2024systematic}.}
  \label{har_batch_removal_compasion}
\end{figure}

We show the plots of the denoising batch effect as below. The results on To326K~\cite{to2024multi} from the Bone tissue are plotted in Fig.~\ref{bone_batch_removal_compasion}. The results on Kuppel139K~\cite{kuppe2022spatial} from the Heart tissue are plotted in Fig.~\ref{heart_batch_removal_compasion}. The results on Morabitol130K~\cite{cortex63k} from the Cortex tissue are plotted in Fig.~\ref{cortex_batch_removal_compasion}. The results on PBMC169K~\cite{de2024systematic} from the PBMC tissue are plotted in Fig.~\ref{har_batch_removal_compasion}. In addition, the results of all metrics are listed in Table~\ref{tab:batch_correction_metrics}.

\begin{table}[H]
    \centering
    \caption{Comparison of Batch Correction and Biological Conservation Metrics Across Datasets. The best performance for each metric within each dataset is highlighted in bold.}
    \label{tab:batch_correction_metrics}
    \resizebox{\textwidth}{!}{%
    \begin{tabular}{C{2.8cm}C{2.5cm}ccccc}
        \toprule
        \textbf{\centering Dataset} & \textbf{\centering Method} & \multicolumn{2}{c}{\textbf{Batch Correction Metrics}} & \multicolumn{3}{c}{\textbf{Biological Conservation Metrics}} \\
        \cmidrule(lr){3-4} \cmidrule(lr){5-7}
        & & $\text{ASW}_{\text{batch}}$ & GraphConn & $\text{ASW}_{\text{cell}}$ & $\text{ARI}_{\text{cell}}$ & $\text{NMI}_{\text{cell}}$ \\
        \midrule
        \multirow{8}{2.8cm}{\centering \textbf{Cortex} Morabito130K~\cite{cortex63k}\\ batch 1/2} 
        & scVI~\cite{scvi} & 0.8863 & 0.8595 & 0.5199 & 0.4537 & 0.4631 \\
        & PeakVI~\cite{peakvi} & 0.9135 & 0.9373 & 0.6022 & 0.8021 & 0.7725 \\
        & PoissonVI~\cite{poissonvi} & 0.9205 & 0.9353 & 0.6021 & 0.7991 & 0.7707 \\
        & Scanorama~\cite{scanorama} & \textbf{0.9391} & 0.9211 & 0.5318 & \textbf{0.8234} & 0.7585 \\
        & Harmony~\cite{harmony} & 0.9174 & 0.9262 & 0.5793 & 0.8032 & 0.7892 \\
        & scANVI~\cite{scanvi} & 0.8790 & 0.9697 & \textbf{0.6585} & 0.6185 & 0.7916 \\
        & Liger~\cite{liger} & 0.8605 & 0.9598 & 0.5689 & 0.7899 & 0.7870 \\
        & SCALEX~\cite{xiong2022online} & 0.9143 & 0.9274 & 0.5791 & 0.7896 & 0.7869 \\
        & scBasset~\cite{yuan2022scbasset} & 0.8052 & 0.9784 & 0.6361 & 0.6876 & 0.7230 \\
        & \textbf{ChromFound} & 0.9323 & \textbf{0.9808} & 0.6408 & 0.7531 & \textbf{0.8380} \\
        \midrule
        \multirow{8}{2.8cm}{\centering \textbf{Bone} To326K~\cite{to2024multi}\\ batch 43/44} 
        & scVI~\cite{scvi} & 0.9073 & 0.9283 & 0.4891 & 0.0070 & 0.0374 \\
        & PeakVI~\cite{peakvi} & 0.9088 & 0.9259 & 0.5145 & 0.1634 & 0.2167 \\
        & PoissonVI~\cite{poissonvi} & \textbf{0.9098} & 0.9259 & 0.5006 & 0.1634 & 0.2167 \\
        & Scanorama~\cite{scanorama} & 0.8836 & 0.9580 & 0.5115 & 0.2325 & 0.3513 \\
        & Harmony~\cite{harmony} & 0.8508 & 0.9544 & 0.5539 & 0.2535 & 0.3066 \\
        & scANVI~\cite{scanvi} & 0.7939 & 0.9506 & 0.5721 & 0.2835 & 0.4604 \\
        & Liger~\cite{liger} & 0.8244 & 0.8874 & 0.5535 & 0.2167 & 0.3668 \\
        & SCALEX~\cite{xiong2022online} & 0.7452 & 0.9506 & 0.5616 & 0.2835 & 0.4604 \\
        & scBasset~\cite{yuan2022scbasset} & 0.5163 & 0.7781 & 0.4339 & 0.0603 & 0.1035 \\
        & \textbf{ChromFound} & 0.8694 & \textbf{0.9885} & \textbf{0.6027} & \textbf{0.6426} & \textbf{0.6772} \\
        \midrule
        \multirow{8}{2.8cm}{\centering \textbf{Heart} Kuppe139K~\cite{kuppe2022spatial}\\ donar av3/av10} 
        & scVI~\cite{scvi} & 0.8128 & 0.8704 & 0.6247 & 0.8186 & 0.8355 \\
        & PeakVI~\cite{peakvi} & 0.9149 & 0.5980 & 0.5215 & 0.6076 & 0.5680 \\
        & PoissonVI~\cite{poissonvi} & \textbf{0.9460} & 0.5980 & 0.5106 & 0.6076 & 0.5680 \\
        & Scanorama~\cite{scanorama} & 0.9378 & 0.7804 & 0.4707 & 0.6869 & 0.7072 \\
        & Harmony~\cite{harmony} & 0.8490 & 0.8407 & 0.5546 & 0.8066 & 0.8298 \\
        & scANVI~\cite{scanvi} & 0.7844 & 0.7715 & 0.4337 & 0.5070 & 0.6553 \\
        & Liger~\cite{liger} & 0.8374 & 0.8407 & 0.4196 & 0.4875 & 0.6287 \\
        & SCALEX~\cite{xiong2022online} & 0.8638 & 0.8164 & 0.5424 & 0.7296 & 0.7711 \\
        & scBasset~\cite{yuan2022scbasset} & 0.7756 & 0.8282 & \textbf{0.6387} & 0.8055 & 0.7417 \\
        & \textbf{ChromFound} & 0.8589 & \textbf{0.8770} & 0.6155 & \textbf{0.9406} & \textbf{0.8980} \\
        \midrule
        \multirow{8}{2.8cm}{\centering \textbf{PBMC} 169K~\cite{de2024systematic}\\ HAR ddseq 1/2} 
        & scVI~\cite{scvi} & 0.7452 & 0.8557 & 0.5794 & 0.5286 & 0.6888 \\
        & PeakVI~\cite{peakvi} & 0.7511 & 0.8266 & 0.6054 & 0.5456 & 0.7005 \\
        & PoissonVI~\cite{poissonvi} & 0.7895 & 0.8266 & 0.5626 & 0.5456 & 0.7005 \\
        & Scanorama~\cite{scanorama} & 0.7964 & 0.8066 & 0.5613 & 0.5456 & 0.7005 \\
        & Harmony~\cite{harmony} & 0.7745 & 0.8120 & 0.5711 & 0.5525 & \textbf{0.7069} \\
        & scANVI~\cite{scanvi} & 0.7480 & 0.8448 & 0.6223 & 0.5114 & 0.6773 \\
        & Liger~\cite{liger} & 0.7755 & 0.8091 & 0.6197 & 0.5481 & 0.7020 \\
        & SCALEX~\cite{xiong2022online} & \textbf{0.8221} & 0.7850 & 0.5534 & \textbf{0.5549} & 0.7003 \\
        & scBasset~\cite{yuan2022scbasset} & 0.7753 & 0.8266 & 0.5695 & 0.5456 & 0.7005 \\
        & \textbf{ChromFound} & 0.7923 & \textbf{0.8511} & \textbf{0.6443} & 0.5730 & 0.7156 \\
        \bottomrule
    \end{tabular}%
    }
\end{table}

\section{Analysis of Cell Type Annotation}
\label{analysis of celltype annotation}
\subsection{Metric Value of Cell Type Annotation}
\label{cell_type_annotation_appendix}
To adapt ChromFound for cell type annotation, we leverage its four-layer pretrained model, extracting the encoder output as a general-purpose representation. The encoder's hidden dimension is expanded from 128 to 256 and projected to a single dimension, forming a pooling tensor. This tensor is then processed by additional multilayer perceptron (MLP) layers to predict cell type logits. During fine-tuning, the pretrained decoder parameters are frozen, while gradients flow through the rest of the backbone to align with the classification objective. Dropout and LayerNorm are applied to mitigate overfitting and stabilize training, respectively. Benchmark methods are implemented using their default parameters as provided in their source codes.

For the experimental setup, we split the training data into 90\% for training and 10\% for validation. Training is conducted over 20 epochs using the AdamW optimizer with an initial learning rate of $5 \times 10^{-4}$ and a 50-step learning rate warmup schedule. The model is selected based on validation performance and evaluated on the test set to compute accuracy and macro F1-score. For certain configurations, such as training on EPF\_hydrop\_2 and testing on VIB\_10xv1\_1, the learning rate is reduced to $2.5 \times 10^{-4}$ to ensure stable convergence. All experiments are performed on a single machine with four NVIDIA A100 GPUs.

\subsection{Confusion Matrix of Cell Type Annotation}
\label{confusion_matrix_appendix}
In the cell classification task where EPF\_hydrop\_3 served as the training set and VIB\_10xv1\_2 as the test set, ChromFound achieves a 4.71\% improvement in accuracy and a 14.69\% increase in macro F1 score compared to the previous SOTA method, CellCano. To further elucidate the specific advancements of ChromFound, we present the confusion matrices of both methods for a more detailed comparative analysis.


Here are some biological insights from the comparison of confusion matrices:

(1) The observed misclassification rate of natural killer cells as cytotoxic T cells by the Cellcano model (19.35\%) and its significant reduction by ChromFound (2.22\%) highlights the importance of accurately distinguishing these functionally related yet distinct immune cell populations. Natural killer cells and cytotoxic T cells share cytotoxic properties, as both can mediate target cell lysis through perforin and granzyme pathways. However, they differ fundamentally in their ontogeny, activation mechanisms, and immune regulation.

(2) The high misclassification rate of CD16+ monocytes as CD14+ T cells by Cellcano (91.91\%), and its substantial reduction by ChromFound (66\%), underscores the challenge of accurately distinguishing myeloid from lymphoid lineages in cell type annotation based on scATAC-seq. CD16+ monocytes, a subset of non-classical monocytes, exhibit distinct chromatin accessibility patterns associated with Fc receptor signaling, inflammatory responses, and patrolling behavior, whereas CD14+ T cells, a less well-characterized subset, retain a predominantly lymphoid epigenetic signature. The improved classification by ChromFound suggests a refined ability to resolve lineage-specific regulatory elements, likely by better capturing differential enhancer accessibility and transcription factor binding landscapes unique to myeloid versus lymphoid cell fate.

(3) Cellcano misclassifies dendritic cells as CD14+ monocytes at a rate of 65.7\% while ChromFound reduces this misclassification to 53.33\%. The misclassification of dendritic cells as CD14+ monocytes is particularly interesting because both cell types share similar transcriptional signatures, especially in immune responses. dendritic cells and monocytes both play critical roles in antigen presentation and inflammation, which may lead to similarities in the chromatin accessibility profiles captured by scATAC-seq. From a biological perspective, this reduction in misclassification also underscores the importance of distinguishing between functionally distinct but phenotypically similar immune cell subsets. Dendritic cells, being key mediators of immune tolerance and initiation, have distinct regulatory networks compared to monocytes, which are more directly involved in inflammatory responses. By refining the accuracy of cell type annotation, ChromFound enables a more precise understanding of immune cell dynamics, particularly in the context of immune responses and disease progression. 
\begin{figure}[htbp]
\vskip 0.2in
\begin{center}
\centerline{\includegraphics[width=0.9\textwidth]{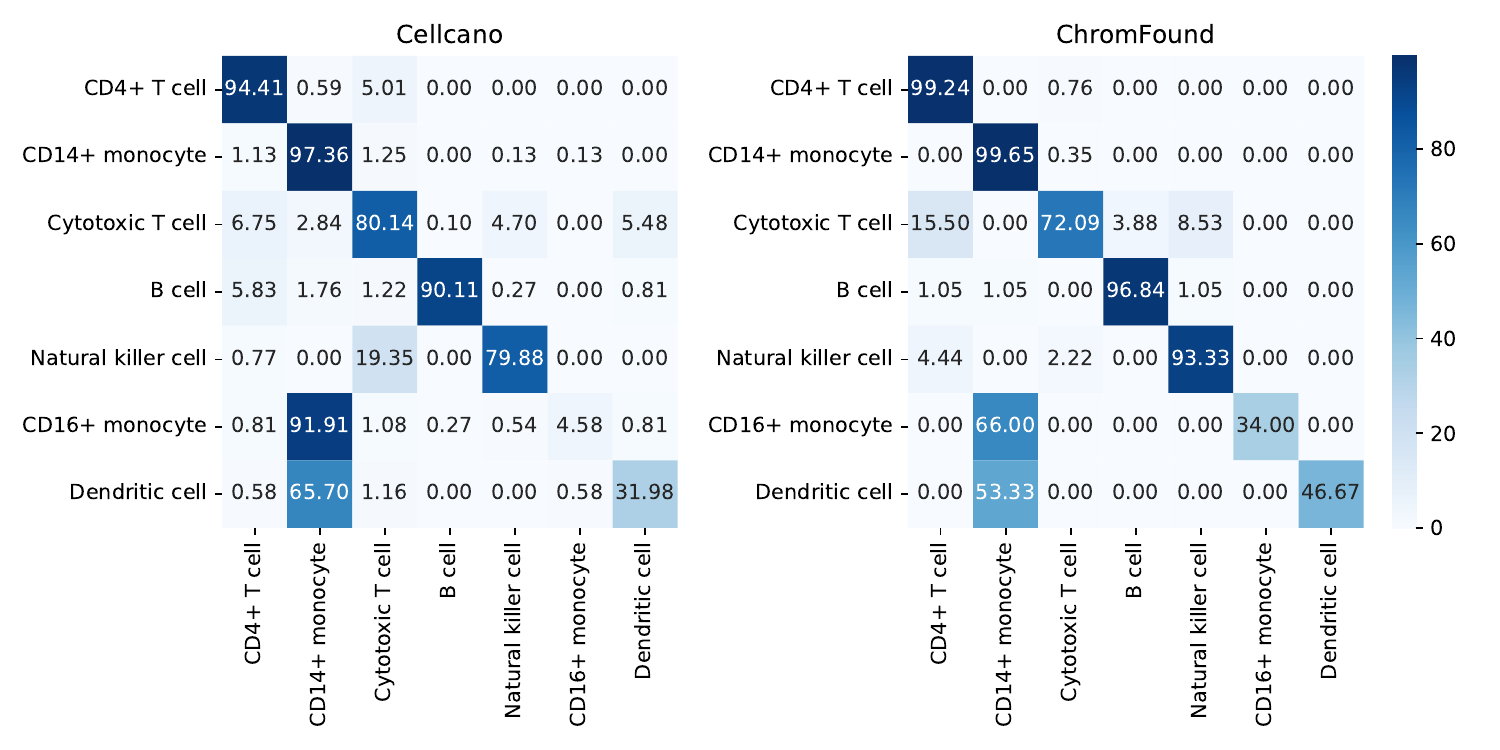}}
\caption{Confusion matrix for cell type annotation task with a comparison of Cellcano and ChromFound.}
\label{confusion_matrix}
\end{center}
\vskip -0.2in
\end{figure}

\section{Benchmark Methods for Downstream Tasks}
\label{benchmark_methods}
This section describes the benchmark methods used for evaluating the performance of ChromFound across various downstream tasks. These methods are selected based on their established effectiveness in single-cell data analysis, spanning cell clustering, cell type annotation, and cross-omics modality prediction.

\subsection{Cell Clustering}
Cell clustering methods aim to group single-cell data into meaningful clusters, often leveraging deep learning and generative models to handle the high dimensionality and sparsity of scATAC-seq data.

\subsubsection{SCALE}
SCALE~\cite{xiong2019scale} is a deep generative framework that employs probabilistic Gaussian Mixture Models to analyze high-dimensional scATAC-seq data. It outperforms existing tools in clustering by effectively capturing latent structures in the data.

\subsubsection{SCALEX}
SCALEX~\cite{xiong2022online} is a deep learning method designed for integrating single-cell data. It projects cells into a batch-invariant embedding space in an online manner, enabling robust cell representation and achieving excellent clustering performance.

\subsubsection{CASTLE}
CASTLE~\cite{cui2024discrete} is a deep generative model tailored for single-cell epigenomic data. It uses a vector-quantized variational autoencoder framework~\cite{van2017neural} to extract discrete latent embeddings, excelling in cell clustering tasks by addressing the challenges of high dimensionality and sparsity.
\subsubsection{scBasset}
scBasset~\cite{yuan2022scbasset} is a sequence-based convolutional neural network method designed for modeling single-cell ATAC-seq data. By utilizing DNA sequence information from accessibility peaks and the expressive power of neural networks, scBasset achieves superior performance in tasks such as cell type identification and data integration across single-cell ATAC-seq and multiome datasets.

\subsubsection{PeakVI}
PeakVI~\cite{peakvi} is a variational autoencoder (VAE)–based probabilistic model for single-cell chromatin accessibility data. It captures both biological and technical variation through a hierarchical generative process, enabling effective batch correction and denoising. The learned latent representation supports diverse downstream tasks such as clustering and differential accessibility analysis, providing a strong probabilistic baseline for scATAC-seq modeling.

\subsubsection{PoissonVI}
PoissonVI~\cite{poissonvi} extends the VAE framework by modeling raw accessibility counts with a Poisson likelihood instead of binarized inputs. This design preserves quantitative accessibility information and improves sensitivity to subtle regulatory differences. PoissonVI achieves robust performance across clustering and integration tasks while maintaining a clear probabilistic interpretation.

\subsubsection{SnapATAC2}
SnapATAC2~\cite{snapatac} is a scalable framework for large-scale scATAC-seq analysis. It represents each cell by a high-dimensional accessibility profile and applies diffusion- and spectral-based dimensionality reduction to reveal chromatin structure. With optimized memory management, parallel I/O, and GPU acceleration, SnapATAC2 efficiently handles datasets with millions of cells and serves as a strong non–deep learning baseline.

\subsubsection{Signac}
Signac~\cite{signac} is an R package built on the Seurat ecosystem for single-cell chromatin accessibility analysis. It integrates preprocessing, feature selection, dimensionality reduction, and visualization within a unified workflow, and supports multimodal integration with scRNA-seq data. Owing to its usability and reproducibility, Signac remains a widely adopted toolkit for exploratory and integrative single-cell epigenomic analysis.

\subsection{Denoising Batch Effect}
Batch effect removal methods aim to align single-cell data across different batches while preserving biological signals, addressing technical variations that can obscure true biological differences. In this evaluation, we also benchmark SCALEX~\cite{xiong2022online}, previously introduced for cell clustering, and scBasset~\cite{yuan2022scbasset}, noted for its sequence-based modeling, alongside other methods. These are included in the comparison for batch effect correction within their respective frameworks, with their detailed definitions provided in prior sections.

\subsubsection{Harmony}
Harmony~\cite{harmony} is a prominent batch correction technique in single-cell analysis, designed to harmonize datasets from diverse experimental batches. It employs an iterative strategy to align cells into a shared low-dimensional embedding, optimizing the similarity of cell types across batches while retaining biological variability, making it highly effective for downstream analyses in single-cell genomics.

\subsubsection{scVI}
scVI~\cite{scvi} is a package for end-to-end analysis of single-cell omics data. The package is composed of several deep generative models for omics data analysis.

\subsubsection{Scanorama}
Scanorama~\cite{scanorama} is designed to be used in scRNA-seq pipelines downstream of noise-reduction methods, including those for imputation and highly-variable gene filtering. The results of Scanorama integration and batch correction can then be used as input to other tools for clustering, visualization, and analysis of scRNA sequences.

\subsubsection{scANVI}
scANVI~\cite{scanvi} (single-cell ANnotation using Variational Inference; Python class SCANVI) is a semi-supervised model for single-cell transcriptomics data. In a sense, it can be seen as a scVI~\cite{scvi} extension that can leverage the cell type knowledge for a subset of the cells present in the data sets to infer the states of the rest of the cells. For this reason, scANVI can help annotate a data set of unlabelled cells from manually annotated atlases.

\subsubsection{Liger}
Liger~\cite{liger} (Linked Inference of Genomic Experimental Relationships) integrates multi-omics single-cell data via integrative non-negative matrix factorization (iNMF) to define shared cell identities across protocols/species.

\subsection{Cell Type Annotation}
Cell type annotation methods focus on supervised learning approaches to assign cells to predefined classes, leveraging advanced neural network architectures for improved accuracy.

\subsubsection{Cellcano}
Cellcano~\cite{ma2023cellcano} utilizes gene expression scores as input and employs a dual-phase training framework to achieve enhanced accuracy in cell type annotation across diverse datasets.

\subsubsection{EpiAnno}
EpiAnno~\cite{chen2022cell} implements a probabilistic generative model with a Bayesian neural network. It delivers remarkable performance in cell type annotation, particularly when applied to diverse single-cell datasets.

\subsection{Cross-omics Prediction}
Cross-omics modality prediction methods aim to infer one modality (e.g., ATAC-seq) from another (e.g., RNA-seq), often using integrative neural network frameworks to model the relationships between modalities.

\subsubsection{BABEL}
BABEL~\cite{wu2021babel} leverages an interoperable neural network model to translate between the transcriptome and chromatin profiles of individual cells, enabling effective cross-omics prediction.

\subsubsection{CMAE}
CMAE~\cite{yang2021multi} learns a probabilistic coupling between different data modalities using autoencoders. It provides a robust framework for integrating and translating between single-cell data modalities.

\subsubsection{scMoGNN}
scMoGNN~\cite{wen2022graph}, an official winner in the overall ranking of modality prediction from the NeurIPS 2021 Competition, presents a general Graph Neural Network framework to facilitate multimodal single-cell data analysis, demonstrating superior performance in cross-omics prediction tasks.

\subsubsection{scMM}
scMM~\cite{minoura2021scmm} is a novel deep generative model-based framework for single-cell multi-omics data analysis (e.g., transcriptome and chromatin accessibility). It employs a mixture-of-experts multimodal approach to extract interpretable joint representations and enable cross-modal generation, achieving end-to-end learning by modeling raw count data with modality-specific probability distributions.

\subsection{Biological Application: Predicting Enhancer-Gene Link and Perturbation Response}
\label{biological_application}
This section outlines the benchmark methods evaluated for the biological application of predicting enhancer-gene links and perturbation responses using ChromFound. This task leverages single-cell data to infer regulatory relationships and responses to perturbations, with a focus on maintaining biological relevance. BABEL~\cite{wu2021babel} and CMAE~\cite{yang2021multi}, the second best models in cross-omics prediction tasks, are included in this application aim to model the regulatory interactions between enhancers and genes, 

\label{pretraing_implementation}

\section{Evaluation Metrics for Downstream Tasks}
\label{metrics}
This appendix details the evaluation metrics used for the downstream tasks in our study, organized by task. Each metric is defined mathematically to ensure clarity and reproducibility, with ranges and performance interpretations provided where applicable.

\subsection{Cell Clustering}
Cell clustering is an unsupervised learning task that groups single-cell data into clusters. When ground-truth labels (e.g., cell types) are available, we evaluate clustering performance using the following metrics.

\subsubsection{Adjusted Rand Index (ARI)}
The Rand Index (RI) measures the proportion of correctly grouped sample pairs:
\[
\text{RI} = \frac{\text{TP} + \text{TN}}{\text{TP} + \text{TN} + \text{FP} + \text{FN}},
\]
where TP, TN, FP, and FN denote the numbers of true positives, true negatives, false positives, and false negatives, respectively, based on whether pairs are correctly grouped together or apart in both the clustering and ground-truth labels. The Adjusted Rand Index (ARI) corrects for chance by adjusting the expected RI under random labeling:
\[
\text{ARI} = \frac{\text{RI} - \mathbb{E}[\text{RI}]}{\max(\text{RI}) - \mathbb{E}[\text{RI}]},
\]
ranging from $-1$ to $1$, where higher values indicate better agreement between clustering results and ground-truth labels.

\subsubsection{Fowlkes-Mallows Index (FMI)}
The Fowlkes-Mallows Index (FMI) combines precision and recall for clustering:
\[
\text{FMI} = \sqrt{\frac{\text{TP}}{\text{TP} + \text{FP}} \cdot \frac{\text{TP}}{\text{TP} + \text{FN}}},
\]
where TP, FP, and FN are defined as above. It ranges from $0$ to $1$, with higher values indicating better clustering performance.

\subsubsection{Normalized Mutual Information (NMI)}
NMI quantifies the shared information between clustering $C$ and ground-truth $Y$:
\[
\text{NMI}(Y, C) = \frac{2 \cdot I(Y; C)}{H(Y) + H(C)},
\]
where $I(\cdot;\cdot)$ is mutual information and $H(\cdot)$ is entropy. It ranges from $0$ to $1$, with higher values indicating greater alignment between the clustering and ground-truth labels.

\subsubsection{Adjusted Mutual Information (AMI)}
AMI adjusts NMI for chance by accounting for expected mutual information under random labeling:
\[
\text{AMI} = \frac{I(Y; C) - \mathbb{E}[I(Y; C)]}{\max(I(Y; C)) - \mathbb{E}[I(Y; C)]},
\]
ranging from $0$ to $1$, where higher values indicate a more robust agreement between clustering and ground-truth, especially in datasets with many clusters.

\subsection{Denoising Batch Effect}
Denoising Batch Effect aligns single-cell data across batches while preserving biological signals. We evaluate performance using biological conservation and batch mixing metrics implemented in~\cite{scib}.

\subsubsection{Biological Conservation Metrics}
We use the following metrics to assess the preservation of biological signals, where $C$ denotes the set of cell types.

\paragraph{Normalized Mutual Information (NMI\textsubscript{cell})}
\[
\text{NMI}_{\text{cell}}(Y, C) = \frac{2 \cdot I(Y; C)}{H(Y) + H(C)},
\]
where $I(\cdot;\cdot)$ is mutual information and $H(\cdot)$ is entropy. It ranges from $0$ to $1$, with higher values indicating better alignment with ground-truth cell types.

\paragraph{Adjusted Rand Index (ARI\textsubscript{cell})}
\[
\text{ARI}_{\text{cell}} = \frac{\text{RI} - \mathbb{E}[\text{RI}]}{\max(\text{RI}) - \mathbb{E}[\text{RI}]},
\]
where $\text{RI} = \frac{\text{TP} + \text{TN}}{\text{TP} + \text{TN} + \text{FP} + \text{FN}}$ and TP, TN, FP, FN are defined as in the clustering section. It ranges from $-1$ to $1$, with higher values indicating better clustering agreement with cell types.

\paragraph{Average Silhouette Width (ASW\textsubscript{cell})}
\[
\text{ASW}_{\text{cell}} = \frac{1}{N} \sum_{i=1}^N \frac{b(i) - a(i)}{\max(a(i), b(i))},
\]
where $a(i)$ is the average distance of cell $i$ to others in its cell type, and $b(i)$ is the smallest average distance to another cell type. It ranges from $-1$ to $1$, with higher values indicating better preservation of biological clustering.

\paragraph{Average Biological Score (AvgBIO)}
\[
\text{AvgBIO} = \frac{\text{ARI}_{\text{cell}} + \text{NMI}_{\text{cell}} + \text{ASW}_{\text{cell}}}{3},
\]
averaging the biological conservation metrics, ranging from $0$ to $1$, with higher values indicating better preservation of biological signals.

\subsubsection{Batch Mixing Metrics}
We assess batch mixing using the following metrics, where $B$ denotes the set of batches.

\paragraph{Inverse Average Silhouette Width (ASW\textsubscript{batch})}
\[
\text{ASW}_{\text{batch}} = 1 - \left| \frac{1}{N} \sum_{i=1}^N \frac{b(i) - a(i)}{\max(a(i), b(i))} \right|,
\]
where $a(i)$ and $b(i)$ are computed with respect to batch labels, with $a(i)$ as the average distance to other cells in the same batch and $b(i)$ as the smallest average distance to cells in another batch. It ranges from $0$ to $1$, with higher values indicating better batch mixing.

\paragraph{Graph Connectivity (GraphConn)}
\[
\text{GraphConn} = \frac{1}{|C|} \sum_{c \in C} \frac{|\text{LCC}(G_c^{\text{kNN}})|}{N_c},
\]
where $\text{LCC}(G_c^{\text{kNN}})$ is the size of the largest connected component in the k-nearest neighbors (kNN) graph of cells in cell type $c$, and $N_c$ is the number of cells in $c$. It ranges from $0$ to $1$, with higher values indicating better connectivity across batches.

\paragraph{Average Batch Score (AvgBATCH)}
\[
\text{AvgBATCH} = \frac{\text{ASW}_{\text{batch}} + \text{GraphConn}}{2},
\]
averaging the batch mixing metrics, ranging from $0$ to $1$, with higher values indicating better batch correction.

\subsection{Cell Type Annotation}
Cell type annotation is a supervised task that assigns cells to ground-truth cell type. We use the following metrics to evaluate performance.

\subsubsection{Accuracy}
Accuracy measures the proportion of correctly classified cells:
\[
\text{Accuracy} = \frac{\sum_{c \in C} \text{tp}_c}{\sum_{c \in C} N_c},
\]
where $\text{tp}_c$ is the number of true positives for cell type $c$, and $N_c$ is the total number of cells in $c$. It ranges from $0$ to $1$, with higher values indicating better classification performance.

\subsubsection{Macro F1 Score}
The Macro F1 Score averages the F1 scores across all cell types:
\[
\text{Macro F1} = \frac{1}{|C|} \sum_{c \in C} F1_c, \quad F1_c = \frac{2 \cdot \text{Precision}_c \cdot \text{Recall}_c}{\text{Precision}_c + \text{Recall}_c},
\]
where $\text{Precision}_c = \frac{\text{tp}_c}{\text{tp}_c + \text{fp}_c}$ and $\text{Recall}_c = \frac{\text{tp}_c}{\text{tp}_c + \text{fn}_c}$, with $\text{fp}_c$ and $\text{fn}_c$ as false positives and false negatives for cell type $c$. It ranges from $0$ to $1$, with higher values indicating better balanced performance across classes.

\subsection{Cross-Omics Prediction}
Cross-omics prediction infers one modality (e.g., ATAC-seq) from another (e.g., RNA-seq). We evaluate performance using correlation-based metrics.

\subsubsection{Pearson Correlation Coefficient (PCC)}
PCC measures the linear correlation between predicted ($x_i$) and true ($y_i$) values:
\[
\text{PCC} = \frac{\sum_{i=1}^N (x_i - \bar{x})(y_i - \bar{y})}{\sqrt{\sum_{i=1}^N (x_i - \bar{x})^2} \cdot \sqrt{\sum_{i=1}^N (y_i - \bar{y})^2}},
\]
where $\bar{x}$ and $\bar{y}$ are the means of the predicted and true values, respectively. It ranges from $-1$ to $1$, with higher (closer to $1$) values indicating stronger positive linear correlation.

\subsubsection{Concordance Correlation Coefficient (CCC)}
CCC extends PCC by accounting for bias and scale differences:
\[
\text{CCC} = \frac{2 \cdot \rho \cdot \sigma_x \cdot \sigma_y}{\sigma_x^2 + \sigma_y^2 + (\mu_x - \mu_y)^2},
\]
where $\rho$ is the PCC, $\sigma_x$ and $\sigma_y$ are the standard deviations of the predicted and true values, and $\mu_x$ and $\mu_y$ are their means. It ranges from $-1$ to $1$, with higher (closer to $1$) values indicating better agreement in both correlation and scale.

\subsection{Predicting Enhancer-Gene Links and Perturbation Response}
These tasks involve predicting regulatory relationships or perturbation outcomes, often treated as classification problems. We use the following metrics.

\subsubsection{Area Under the Receiver Operating Characteristic Curve (AUC-ROC)}
The Receiver Operating Characteristic (ROC) curve plots the True Positive Rate (TPR) against the False Positive Rate (FPR) at various classification thresholds:
\[
\text{TPR} = \frac{\text{TP}}{\text{TP} + \text{FN}}, \quad \text{FPR} = \frac{\text{FP}}{\text{FP} + \text{TN}},
\]
where TP, TN, FP, and FN are true positives, true negatives, false positives, and false negatives, respectively. The Area Under the ROC Curve (AUC-ROC) quantifies the overall performance across all thresholds, ranging from $0$ to $1$, with higher values indicating better classification performance, where $1$ represents perfect classification and $0.5$ represents random guessing.

\end{document}